\documentstyle[a4,12pt]{article}
\renewcommand{\baselinestretch}{1.5}
\newtheorem{df}{DEFINITION}
\newtheorem{th}{THEOREM}
\newtheorem{fact}{FACT}

\begin{document}
\def\SL{$\widetilde{SL}(2,R)$}
\def\refmark#1{{[#1]}}

\def\abstract#1{\begin{center}{\large ABSTRACT}\end{center} \par #1}
\def\title#1{\begin{center}{\large #1}\end{center}}
\def\author#1{\begin{center}{\sc #1}\end{center}}
\def\address#1{\begin{center}{\it #1}\end{center}}
\def\pubnum{208/COSMO-26}  
\begin{titlepage}
\hfill
\parbox{6cm}{{TIT/HEP-\pubnum} \par {Revised version}}
\par
\vspace{7mm}
\title{Compact Homogeneous Universes}
\vskip 1cm
\author{ Tatsuhiko KOIKE\footnote{E-mail address:
bartok@phys.titech.ac.jp},
Masayuki TANIMOTO \footnote{E-mail address: prince@phys.titech.ac.jp}
and Akio HOSOYA\footnote{E-mail address: ahosoya@phys.titech.ac.jp}}
\address{Department of Physics, Tokyo Institute of \\ Technology,
Oh-Okayama, Meguro-ku, Tokyo 152, Japan}
\vskip 1 cm
\abstract{
A thorough classification of the topologies of compact
homogeneous universes is given except for the hyperbolic
spaces, and  their global degrees
of freedom are completely worked out.
To obtain compact universes,  spatial points are identified
by discrete subgroups of the isometry
group of the generalized Thurston geometries, which are related
to the Bianchi and the Kantowski-Sachs-Nariai universes.
Corresponding to this procedure their total degrees of freedom are
shown to
be categorised into those of the universal covering space
and the Teichm\"uller parameters.
The former are given by constructing homogeneous metrics on simply
connected
manifold. The Teichm\"uller spaces are also given
by explicitly constructing expressions for the discrete subgroups of
the
isometry group.
}
\vfill
PACS number(s): 04.20.-q, 04.20.Gz, 02.40.-k.

{\em to appear in Journal of Mathematical Physics}
\end{titlepage}

\pagenumbering{arabic}
\setcounter{page}{2}
\newcommand{\babst}{\begin{abstract}}
\newcommand{\eabst}{\end{abstract}}
\newcommand{\bdf}{\begin{df}}
\newcommand{\edf}{\end{df}}
\newcommand{\bth}{\begin{th}}
\newcommand{\eth}{\end{th}}
\newcommand{\bcenter}{\begin{center}}
\newcommand{\ecenter}{\end{center}}
\newcommand{\beq}{\begin{equation}}
\newcommand{\eeq}{\end{equation}}
\newcommand{\bea}{\begin{eqnarray}}
\newcommand{\beqa}{\begin{eqnarray}}
\newcommand{\beqna}{\begin{eqnarray}}
\newcommand{\eea}{\end{eqnarray}}
\newcommand{\eeqa}{\end{eqnarray}}
\newcommand{\eeqna}{\end{eqnarray}}
\newcommand{\ba}{\begin{array}}
\newcommand{\ea}{\end{array}}
\newcommand{\btabu}{\begin{tabular}}
\newcommand{\etabu}{\end{tabular}}
\newcommand{\btable}{\begin{table}}
\newcommand{\etable}{\end{table}}
\newcommand{\bthebib}{\begin{thebibliography}}
\newcommand{\ethebib}{\end{thebibliography}}
\newcommand{\nono}{\nonumber}
\newcommand{\nonum}{\nonumber}

\newcommand{\compulsoryindent}{\hspace*\parindent}

\def\sect#1{\section{#1}\compulsoryindent}
\def\subsect#1{\subsection{#1}\compulsoryindent}
\def\subsubsect#1{\subsubsection{#1}\compulsoryindent}

\def\mtc#1#2#3{\multicolumn{#1}{#2}{#3}}

\def\eg{{\it e.g.\ }}
\def\ie{{\it i.e.\ }}
\def\cf{{\it cf\/\ }}
\def\etc{{\it etc}}
\def\etal{{\it et al\/\ }}


\def\ti{\tilde}
\def\wti{\widetilde}
\def\pr{^\prime}
\def\lsp#1#2{{}^{(#1)}\!#2}
\def\abs#1{\left | #1 \right |}
\def\rcp#1{{1\over #1}}
\def\paren#1{\left( #1 \right)}
\def\brace#1{\left\{ #1 \right\}}
\def\bra#1{\left[ #1 \right]}
\def\angl#1{\left\langle #1 \right\rangle}
\def\Re{\mbox{\rm Re}}
\def\Im{\mbox{\rm Im}}
\def\bm{\boldmath}
\def\any{\forall}

\def\pd#1#2{{\partial#1\over\partial#2}}
\def\cb#1{{\partial \over \partial #1}}
\def\^{\wedge}

\def\al{alpha}
\def\bt{\beta}
\def\gm{\gamma}
\def\dl{\delta}
\def\ep{\epsilon}
\def\vep{\varepsilon}
\def\zt{\zeta}
\def\et{\eta}
\def\tht{\theta}
\def\vth{\vartheta}
\def\lm{\lambda}
\def\sig{\sigma}
\def\om{\omega}

\def\N{\mbox{\bf N}}
\def\R{\mbox{\bf R}}
\def\C{\mbox{\bf C}}
\def\Z{\mbox{\bf Z}}
\def\ZZ#1{{\bf Z}_{#1}}
\def\E#1{E^#1}
\def\S#1{S^#1}
\def\T#1{T^#1}
\def\P#1{P^#1}
\def\H#1{H^#1}
\def\O#1{\mbox{\rm O($#1$)}}
\def\IO#1{\mbox{\rm IO($#1$)}}
\def\SO#1{\mbox{\rm SO($#1$)}}
\def\ISO#1{\mbox{\rm ISO($#1$)}}
\def\U#1{\mbox{\rm U($#1$)}}
\def\SU#1{\mbox{\rm SU($#1$)}}
\def\GLR#1{\mbox{\rm GL($#1$,\R)}}
\def\SLR#1{\mbox{\rm SL($#1$,\R)}}
\def\PSLR#1{\mbox{\rm PSL($#1$,\R)}}
\def\GLC#1{\mbox{\rm GL($#1$,\C)}}
\def\SLC#1{\mbox{\rm SL($#1$,\C)}}
\def\PSLC#1{\mbox{\rm PSL($#1$,\C)}}
\def\Sol{\mbox{\rm Sol}}
\def\Nil{\mbox{\rm Nil}}
\def\omatrix{\matrix 0 0 0 0}

\def\x{\bf x}
\def\dif#1{\pd{}#1}
\def\Del#1#2{\pd{#1}{#2}}

\def\SP{\hspace{8 mm}}
\def\Teich{{Teichm\"{u}ller }}
\def\Poin{{Poincar\'{e} }}
\def\Gam{\mbox{$\Gamma$} }

\def\a{{\bf a}}
\def\Isom{{\rm Isom}}
\def\tiSLR#1{\mbox{$\wti{\rm SL}({#1},\R)$}}

\def\lam{\lambda}
\def\aaa{e^{2 \lam}}
\def\aam{e^{-2 \lam}}
\def\bbb{e^{(2 / \sqrt{3}) \lam}}
\def\bbm{e^{-(4 / \sqrt{3}) \lam}}
\def\bb{e^{2(\lam_+ / \sqrt{3}+\lam_-)}}
\def\cc{e^{2(\lam_+ / \sqrt{3}-\lam_-)}}
\def\dd{e^{-(4 / \sqrt{3}) \lam_+}}
\def\semi{\mbox{$\Join$}}
\def\mmp#1#2#3{\mbox{$(#1,#2,#3) \mapsto (-#1,-#2,#3)$}}
\def\mpp#1#2#3{\mbox{$(#1,#2,#3) \mapsto (-#1,#2,#3)$}}
\def\mpm#1#2#3{\mbox{$(#1,#2,#3) \mapsto (-#1,#2,-#3)$}}
\def\pmm#1#2#3{\mbox{$(#1,#2,#3) \mapsto (#1,-#2,-#3)$}}
\def\rppm#1#2#3{\mbox{$(#1,#2,#3) \mapsto (#2,#1,-#3)$}}
\def\rmmm#1#2#3{\mbox{$(#1,#2,#3) \mapsto (-#2,-#1,-#3)$}}
\def\pmpmp#1#2#3{\mbox{$(#1,#2,#3) \mapsto (\pm #1,\pm #2,#3)$}}
\def\pmppm#1#2#3{\mbox{$(#1,#2,#3) \mapsto (\pm #1, #2,\pm #3)$}}
\def\ppmpm#1#2#3{\mbox{$(#1,#2,#3) \mapsto (#1,\pm #2,\pm#3)$}}
\def\rpmpmm#1#2#3{\mbox{$(#1,#2,#3) \mapsto (\pm #2,\pm #1,-#3)$}}
\def\rrmmm#1#2#3{\mbox{$(#1,#2,#3) \mapsto (-#3,-#2,-#1)$}}

\def\wa{\!\!\!\!&=&\!\!\!\!}
\def\wb{\!\!\!\!&\equiv &\!\!\!\!}
\def\lvector#1#2#3#4{\paren{
    \begin{array}{c} #1 \\ #2 \\ #3 \\ #4 \end{array}}}
\def\vector#1#2#3{\paren{\begin{array}{c} #1 \\ #2 \\ #3 \end{array}}}
\def\svector#1#2{\paren{\begin{array}{c} #1 \\ #2 \end{array}}}
\def\matrix#1#2#3#4#5#6#7#8#9{
    \left( \begin{array}{ccc}
            #1 & #2 & #3 \\ #4 & #5 & #6 \\ #7 & #8 & #9
    \end{array} \right) }
\def\smatrix#1#2#3#4{
    \left( \begin{array}{cc} #1 & #2 \\ #3 & #4 \end{array} \right) }
\def\manbo#1{}  

\def\Teich{{Teichm\"{u}ller }}
\def\Poin{{Poincar\'{e} }}
\def\Gam{ \mbox{$\Gamma$} }
\def\a{{\bf a}}
\def\G{{\rm\bf G}}
\def\GT{{\rm\bf\tilde{G}}}
\def\rh{h^{1/2}}
\def\d{{\rm d}}
\def\Bsvn{VII${}_0$ }
\def\ho{\overline{h}}
\def\eo{\overline{{\rm e}}}
\def\ao{\overline{\alpha}}
\def\tR{{}^{(3)}\! R}
\def\fR{{}^{(4)}\! R}
\def\SR{\mbox{$S^2\times E^1$} }
\def\Isom{{\rm Isom}}
\def\SL{\widetilde{{\rm SL}}(2,\R)}
\def\SLx{{\rm SL}(2,\R)}
\def\Nil{{\rm Nil}}
\def\Sol{{\rm Sol}}
\def\o{{\bf O}} 

\def\univ{\mbox{$\tilde{M}$} }
\def\Tp{\mbox{\boldmath $\tau$} }
\def\Tpp{{\tau} }
\def\rep{ \mbox{$\dh a b ^{{\rm (rep)}}$} }
\def\shifth{ \mbox{$\phi_{\Tpp *}\rep$} }
\def\conf{e^{2\alpha} }
\def\stt{\sin^2\!\theta}
\def\eqmap{\overline{\phi}}
\def\Rep{{\rm Rep}}
\def\teich{{\rm Teich}}
\def\univh{\mbox{$\tilde{h}_{ab}$} }

\def\e#1{{\rm\bf e}_{#1}}
\def\ue#1{{\rm\bf e}^{#1}}
\def\cG#1{\mbox{${\cal G}_#1$}}
\def\uh#1#2{h^{#1#2}}
\def\dh#1#2{h_{#1#2}}
\def\ug#1#2{g^{#1#2}}
\def\dg#1#2{g_{#1#2}}
\def\dB#1#2{B_{#1#2}}
\def\uB#1#2{B^{#1#2}}
\def\dBa#1#2{{\overline{B}_{#1#2}}}
\def\uBa#1#2{{\overline{B}^{#1#2}}}
\def\dha#1#2{{h_{(#1)(#2)}}}
\def\uha#1#2{{h^{(#1)(#2)}}}
\def\ee#1#2{{\rm e}^{(#1)}_{#2}}
\def\BVII#1{\mbox{Bianchi VII${}_#1$} }
\def\BVI#1{\mbox{Bianchi VI${}_#1$} }
\def\four#1{{}^{(4)}\! #1}

\def\AS{Ashtekar and Samuel }
\def\goes{\rightarrow}


\sect{Introduction}
 In the standard cosmology we normally assume homogeneity
and isotropy of the universe.  On the basis of this cosmological
principle
the Friedmann-Robertson-Walker metric is constructed and Big-Bang
Cosmology
has been successfully developed in the theory of Einstein gravity
\cite{Weinberg}.  The homogeneous but {\em anisotropic} universe
models, the
Bianchi \cite{LL,Wa} and the Kantowski-Sachs-Nariai universes
\cite{Na,KS},
have been studied from various motivations \cite{BKL}, many of which
contribute to grasp aspects of more general and more complicated
situations
in general relativity or in cosmology.  For example, people have
attempted to
clarify the properties of the initial singularity and also attacked
the basic
problem why the present universe looks so isotropic by studying the
time
evolution of the homogeneous anisotropic models \cite{Ryan}.  The
investigation of {\em compact} universes may be important when we are
interested in the possibility of our universe to have a non-trivial
topology
or its topology change.  The former aspect was studied by Fang and
Sato
\cite{FS} and by Ellis \cite{El} in a special model of torus
universe.  For
the latter we note that the famous theorem by Geroch \cite{Geroch} on
topology change assumes the compactness of spatial manifold.  It will
be more
intriguing to study the geometries of compact (locally) homogeneous
universes
when we investigate minisuperspace models in quantum gravity.  There
it is
necessary to make explicit the global degrees of freedom
corresponding to
deformation of the compact universes.

For such investigations, we need a classification of compact
homogeneous
spaces having nontrivial topologies, which are not considered in the
traditional Bianchi classification.
Ashtekar and Samuel \cite{AS} investigated global degrees of freedom
of compact locally homogeneous universes.
However, it seems to us that their definition of locally homogeneous
spaces is so restrictive that many possible types of compact universes
are excluded.
 For example, in the Bianchi V model, they have excluded the compact
hyperbolic spaces
by restricting the identifications used in the compactification
to those in the 3-dimensional Bianchi V group.
In fact, the Bianchi V model {\em can} be compactified by using the
largest
6-dimensional isometry group $\mbox{PSL}(2,\bf C) \supset$ Bianchi V
group.
Our definition and treatment of locally homogeneous spaces,
which is overviewed below, may be more natural and
general and coincides with the one adopted by mathematicians.

In a modern point of view, the term ``geometry'' is used in a specific
sense. It is a pair of a manifold $X$ and a group $G$ acting on $X$
transitively with compact point stabilizer (i.e., the isotropy
subgroup
of $G$ at any point of $X$ is compact).
A geometry $(X,G)$ is an alternative expression of a
homogeneous Riemannian manifold $(X,\dg ab)$,
since we can always construct the $G$-invariant metric $\dg ab$ in an
appropriate manner.
Thurston \cite{Th}
(see also \cite{Sc})
proved that any {\it maximal} and simply connected 3-dimensional
geometry which admits a compact quotient is equivalent to one of the
so-called
Thurston eight geometries.
Here, the term ``maximal'' is defined with respect to inclusion
of the group $G$.
In other words, maximal geometries are the spaces having such a large
symmetry that no extra symmetry can be added.
The maximality, however, restricts possible universal covers of
compact
locally homogeneous spaces.
By extending the Thurston's results to {\em nonmaximal} geometries,
we shall show a physically relevant theorem,
and give the explicit forms of the universal cover metrics.
We emphasize that {\it all} three-dimensional compact
homogeneous
manifolds can be obtained from the universal covers with
the metrics given in the generalized theorem.
We call the parameters in the metrics the {\it universal cover
parameters}.

Compact homogeneous spaces can be obtained, as already inferred,
by identifying the points of the universal covers which are
mapped to one another
by the action of an appropriate discrete subgroup of the isometry
group.
This may be well described by the following projection
\beq
\pi:\; \univ\goes M=\univ/\Gam, \label{1-projection}
\eeq
where \univ is a universal covering space, $\Gam$ the discrete
isometry group
and $M$ the compact homogeneous space expressed by the projection
$\pi$.
It is worth noting that $\Gam$ is isomorphic to the
fundamental group of $M$, $\pi_1(M)$.
The {\it \Teich deformations} are expressed by smooth deformations of
$\Gam$.
Hence, it is evident that the \Teich deformations preserve
local quantities such as curvature,
whereas the deformations of a universal cover metric alter them.
We shall present the possible expressions of $\Gam$ explicitly.
We call the parameters in $\Gam$ the {\it \Teich parameters}.

The total (geometrical) degrees of freedom of a compact locally
homogeneous space is defined as the number of the universal cover
parameters plus the number of the \Teich parameters.
For example, we shall see that Bianchi IX universe has two degrees
of freedom of the universal cover and no Teichm\"uller deformations,
and that Bianchi VI(0) universe has one
universal cover parameter and one Teichm\"uller parameter.

In this paper, we will not discuss the dynamics of the compact spaces.
The Einstein gravity will be investigated for the compact locally
homogeneous universes in a separate paper.

The organization of the present paper is the following.
In section 2, some mathematical preliminaries are presented to clearly
define compact locally homogeneous spaces and
the Teichm\"uller deformation.
We also explain Thurston's theorem which plays a central role in
the present work and the eight geometries
which appear in the theorem. The homogeneity preserving diffeomorphism
is also defined.
In section 3, we describe the definition of the compact locally
homogeneous cosmological models
in terms of the mathematical terminologies explained in section 2.
In section 4, we give a complete classification of the types of
universal
covers $\tilde M$. The compact quotients $M= {\tilde M/\Gamma}$
are explicitly constructed in section 5, except for the
compact hyperbolic 3-geometries.
The final section is devoted to conclusion and discussions.

 In the Appendix we present a proof that shows why Bianchi Class B
geometries and $({\bf R}^3,\widetilde{\mbox{SL}}(2,{\bf R}))$  do not
admit
compact quotients if one uses the Bianchi groups to find the discrete
subgroup $\Gamma$.

\sect{Mathematical preliminaries}
 In this section we explain briefly and without proofs
the mathematical terms and facts
which we will use in what follows.
 In Sec.\ref{lochom} we give the definition and the feature of
locally homogeneous manifolds.
 In Sec.\ref{geom} we explain the theorem of Thurston which will play
a great role in this paper.
 In Sec.\ref{teich} we give the definition of the \Teich space.
 In Sec.\ref{eight} we briefly explain each of the Thurston
geometries.
The proofs and more details may be found in Wolf \cite{Wo}
and Scott \cite{Sc}.

\subsect{Locally homogeneous spaces}
\label{lochom}
 A metric on a manifold $M$ is said to be ({\it globally})\/ {\it
homogeneous}
\/ if the isometry group acts transitively on $M$, i.e. for any
points
$p,q\in M$ there exists an isometry which takes $p$ into $q$.
 A metric on a manifold $M$ is said to be {\it locally homogeneous}\/
if for any points $p,q\in M$ there exist neighborhoods $U,V$ and an
isometry
$(U,p)\to (V,q)$.
 The difference between global and local homogeneity is that in the
latter
case the isometry may not be globally defined.
\medskip

 Any locally homogeneous manifold is diffeomorphic to the quotient of
a
simply connected homogeneous manifold by a discrete subgroup of the
isometry group.
 Let $M$ be an arcwise connected compact manifold
with a complete locally homogeneous metric.
Any arcwise connected manifold $M$ has a unique universal covering
manifold
$\ti{M}$ up to diffeomorphisms.
 $\ti{M}$ has a natural metric inherited from $M$ which is the
pullback
of the metric on $M$ by the projection $\ti{M}\to M$. This defines
a complete locally homogeneous metric on $\ti{M}$.
 It is a theorem of Singer \cite{Si}
that a complete locally homogeneous metric on a
simply connected manifold is homogeneous.
 Therefore the metric on $\ti{M}$ is homogeneous and $\ti{M}$
is diffeomorphic to $\Isom\ti{M}/K$ where $\Isom\ti{M}$ is the
isometry
group of $\ti{M}$ and $K$ is the stabiliser of a point in $\ti{M}$.
 $K$ is a compact subgroup of $\Isom\ti{M}$.
 Let $\Gamma$ be the group of covering transformations, which is
isomorphic to the fundamental group $\pi_1 (M)$ of $M$.
 An element of $\Gamma$ is obviously an isometry.
 It follows that $\Gamma$ is a subgroup of $\Isom\ti{M}$.
 Thus $M$ is realized as $\ti{M}/\Gamma$.
 We restrict our attention to orientable manifolds.
 Then each element of $\Gamma$ must preserve the orientation
 of $\ti{M}$.
 This means that in order to get orientable manifolds one has only to
replace
$\Isom\ti{M}$ in the above discussion by $\Isom^+ \ti{M}$, the
orientation
preserving subgroup of $\Isom\ti{M}$.

\medskip

 Conversely, a quotient space of a homogeneous Riemannian manifold by
an
appropriate subgroup of the isometry group is locally homogeneous.
Let $X$ be a manifold with a complete
 homogeneous metric, $\Isom X$ be its isometry group, $K$ be a
stabiliser
 of a point, and $\Gamma$ be a subgroup of $\Isom X$.
 $X/\Gamma$ is Hausdorff if $\Gamma$ acts {\it properly
discontinuously}\/ on
 $X$, i.e. for any compact subset $C$ of $X$,
 $\{\gamma\in \Gamma\;|\;\gamma C\cap C\neq \emptyset \}$ is finite.
 $\Gamma$ acts properly discontinuously on $X$ if and only if $\Gamma$
 is a discrete subgroup of $\Isom X$.
 $X/\Gamma$ is a Riemannian manifold if and only if $\Gamma$
 acts $freely$ on $X$, i.e. the stabilizer of $\Gamma$ is trivial.
 If $\Gamma$ does not necessarily act freely then $X/\Gamma$
 is an {\it orbifold}, i.e. a Hausdorff, paracompact space which is
locally
 homeomorphic to the quotient space of ${\rm R}^n$ by a finite group
 action.
 Evidently $X/\Gamma$ is locally homogeneous if it is a
Riemannian manifold.
 As a result, $X/\Gamma$ is a locally homogeneous Riemannian
manifold if $\Gamma$ is a discrete subgroup of $\Isom X$
and acts freely on $X$.
 So our task is to find out all possible discrete subgroups
 $\Gamma$ of $\Isom X$ which acts freely on
 $X$ and makes $X/\Gamma$ compact.

\subsect{Geometries and Thurston's theorem}
\label{geom}
 In order to state precisely and take advantage of the result
obtained in
 the recent study in 3-manifolds,
 we use the term `geometry' in a specific sense.
 A {\it geometry}\/ is a pair $(X,G)$ where $X$ is a manifold and $G$
 is a group acting transitively on $X$ with compact point
 stabilizers.
 Geometries $(X,G)$ and $(X^\prime,G^\prime)$ are
{\it equivalent}\/ if there is  a diffeomorphism $\phi:X\to X^\prime$
with
$\bar{\phi}:g \mapsto \phi\circ g\circ\phi^{-1}$ being an isomorphism
from $G$ onto $G^\prime$.
 Let us call $\phi$ an {\it equivalence map}.
 A geometry $(X,G^\prime)$ is a {\it subgeometry} of $(X,G)$ if
$G^\prime$ is
 a subgroup of $G$.
 A geometry is {\it maximal}\/ if it is not a proper subgeometry of
any
geometries.
\medskip


 Although a geometry $(X,G)$ is merely a pair of a manifold $X$ and
a group $G$ and has nothing to do with Riemannian metrics on $X$
itself,
one can discuss homogeneous Riemannian metrics from the viewpoint of
geometries.
 This is because one can construct a complete
homogeneous metric from a geometry $(X,G)$ as follows.
 Give a random positive definite quadratic form on the tangent space
$T_p$ of a point $p$ of $X$.
 Average it on $p$ by the invariant volume form of the compact
stabilizer
$G_p$, having a $G_p$-invariant inner
product on $T_p$.
 Carry the quadratic form to each point in $X$ by the action of $G$
which acts transitively on $X$, having the $G$-invariant metric on
$X$.
 Note that there may be many possible Riemannian metrics on $X$ which
correspond to a geometry $(X,G)$.
 Note also that $G$ is not necessarily the isometry group itself but
a subgroup of it.

 We now quote the theorem of Thurston which is of great use
in classifying all compact, locally homogeneous spaces.
\medskip

\begin{th}
\label{th:thurston}
 {\rm(Thurston \cite{Th})}  Any maximal, simply connected
3-dimensional
geometry which admits a compact quotient is equivalent to
the geometry \/$(X,\Isom X)$  where $X$ is one
of \/ $\E3$, $\H3$, $\S3$, $\S2 \times \R$, $\H2 \times \R$,
$\tiSLR2$, \Nil, or \/ \Sol.
\end{th}

\medskip
\noindent
 A brief proof can be found in Scott \cite{Sc}.
 Each $X$ in the theorem is a manifold with a certain `standard'
Riemannian
metric on it.
 Note that this Riemannian metric on $X$ is not of essential
importance in
the theorem but is simply useful to express a group $G$ in a geometry
$(X,G)$ as $\Isom X$ and the action of $G$ on $X$.
 For example, $(\E3,\Isom\E3)$ means the same as $(\R^3,\IO3)$.
 A brief explanation of all geometries is given in Sec.\ref{eight}.

\medskip
 One important feature of the Thurston geometries is that the compact
quotients modeled on the six of them apart from $\H3$ and $\Sol$
admit Seifert bundle structures.
 A Seifert bundle structure is an extended idea of an $\S1$-bundle
structure over a 2-surface.
 Though Seifert bundle itself is a 3-manifold,
the `base space' may be an orbifold
and the fiber at a singular point of the base (critical fiber)
does not have to continue smoothly to those of neighboring points
(regular fibers).
 The singular points of the base are always cone points if the total
space
is orientable.
 An orientable Seifert bundle is specified by the base orbifold,
$\S1$-bundle over a 2-manifold obtained by removing cone points from
the base
orbifold,
Seifert invariants $(\alpha,\beta)$ of each critical fiber,
and the ``Euler number'' $e$.
 The Seifert invariants $(\alpha,\beta)$ are integers and
determine the relation of
a critical fiber and its neighboring fibers; $\alpha$ turns along
the critical fiber is equal to $-\beta$ turns along the regular
fiber.
 The Euler number $e$ is an integer which tells the nontriviality
of the bundle.
  Regarding a manifold as a Seifert bundle makes it easier to
investigate it geometrically, especially in considering possible
deformations of it.

\subsect{The Teichm\"{u}ller space}
\label{teich}
 We give the definition of the \Teich space of a locally homogeneous
manifold.
 Let $\ti{M}$ be a simply connected homogeneous Riemannian manifold
with
its metric $\ti{h}_{ab}$ fixed.
 Let $M$ be a quotient of $\ti{M}$ by a discrete subgroup $\Gamma$ of
$\Isom\ti{M}$.
 The \Teich space of $M$ is the space of all smooth deformations
of $M$ irrespective of its size,
keeping the condition that $M$ is a quotient of
$(\ti{M},\ti{h}_{ab})$.

 Let us put it more precisely.
 Let Rep($M$) denote the space of all discrete and faithful
representations $\rho:\pi_1(M)\to\Isom^+ \ti M$.
 Let us call a diffeomorphism $\phi:\ti{M}\to\ti{M}$ a global
conformal isometry if $\phi_* \ti h_{ab}=\mbox{const.} \cdot \ti
h_{ab}$.
 Let us define a relation $\sim$ in Rep$(M)$
such that $\rho \sim \rho\pr$ holds if there exists
a global conformal isometry $\phi$ of $\ti{M}$ connected to the
identity
which gives $\rho\pr(a)=\phi\circ\rho(a)\circ\phi^{-1}$
for any $a\in\pi_1(M)$.
 It defines an equivalence relation in Rep($M$).
We define the {\it \Teich space}\/ as
\beq
 \mbox{Teich}(M)=\mbox{Rep}(M)/\sim.
\eeq
 The \Teich space is a manifold.
 The numbers used to parametrize the \Teich space are called the
{\it Teichm\"{u}ller parameters}.

 If two representations $\rho$ and $\rho\pr$ correspond to the same
point in Teich$(M)$, the above global conformal isometry $\phi$ on
$\ti M$
induces a well-defined global conformal isometry from
$\ti M/\Gamma$ onto $\ti M/\Gamma\pr$,
where $\Gamma=\rho(\pi_1(M))$ and $\Gamma\pr=\rho\pr(\pi_1(M))$.
 This is because for any $\gamma\in\Gamma$ there exists unique
$\gamma\pr\in\Gamma\pr$ such that
$\phi\circ\gamma=\gamma\pr\circ\phi$,
which guarantees that $\phi(\ti M/\Gamma)=\ti M/\Gamma\pr$.

 We will see in Sec.\ref{univ:results} that homogeneous manifolds
$\ti{M}$ except for the
flat space have no global conformal isometries other than isometries.

 In this case, the \Teich space is Rep$(M)$ modulo conjugation by
$\Isom_0\ti M$.
 In the case of the flat space, the \Teich space is Rep$(M)$
modulo conjugation by $\Isom_0\ti M$ and modulo total size.

\medskip
 The Teichm\"{u}ller parameters constitute a subset of
dynamical variables of locally homogeneous universes.

\subsect{The Thurston geometries}
\label{eight}
 In this section we explain the metric and the isometry group
of the eight Thurston geometries.

\subsubsect{$\E3$}
\label{eight:e3}
$\E3$ is the 3-dimensional flat Riemannian manifold.
 The standard metric is
\begin{equation}
ds^2=dx^2+dy^2+dz^2
\end{equation}
and an isometry is expressed as
\begin{equation}
g(\x)=R\x + \a
\end{equation}
where $R$ is a $3 \times 3$ orthogonal matrix and $\a$ is a constant
vector.
 The above transformations form a group and we denote it by \IO3.
 To preserve the orientation $R$ must be a rotation
matrix, and the corresponding transformations form a subgroup, \ISO3,
of \IO3.

 The generators of $\IO3$ or $\ISO3$, i.e. Killing vectors of $\E3$,
are the following:
\begin{eqnarray}
\label{eq:iso3}
K_1&=&\dif x,            \nonumber\\
K_2&=&\dif y,            \nonumber\\
K_3&=&\dif z,            \nonumber\\
K_4&=&-z \dif y+y\dif z, \nonumber\\
K_5&=&-x \dif z+z\dif x, \nonumber\\
K_6&=&-y \dif x+x\dif y.
\end{eqnarray}
 The first three vector fields represent infinitesimal translations
while the last three represent infinitesimal rotations.

\subsubsect{\Nil}
\label{eight:nil}
 $\Nil$ is the group of the matrices of the form
$\left(
\begin{array}{ccc}
1 & x & z\\
0 & 1 & y\\
0 & 0 & 1
\end{array}
\right)$
with ordinary matrix multiplication.
 In other words, the multiplication is defined by
\beq
(a,b,c)(x,y,z)=(a+x,b+y,c+z+ay).
\eeq
 The standard metric is one of left-invariant metrics.
 A basis of left-invariant 1-forms is
\beq
\sig^1=dz-xdy,\sig^2=dx,\sig^3=dy.
\eeq
 The standard metric is
\beqa
ds^2&=&(\sig^1)^2+(\sig^2)^2+(\sig^3)^2  \nono\\
    &=&dx^2+dy^2+(dz-xdy)^2.
\eeqa
 The isometry group of course has $\Nil$ as its subgroup.
 There is an additional 1-parameter family of isometries isomorphic
to U(1).
 It is written as
\beq
s_\theta:\left( \ba{c}x\\y\\z \ea\right)
\mapsto
\left( \ba{c} R_\theta \left(\ba{c}x\\y\ea\right)\\
z+{1\over2}((x^2-y^2)\cos\theta-2xy\sin\theta)\sin\theta\ea
\right)
\ \ (0\leq\theta<2\pi)
\eeq
where $R_\theta$ is the 2-dimensional rotation matrix of angle
$\theta$.
 Therefore the isometry group is 4-dimensional.
 The isometry group has two components.
 A discrete isometry is given by
\beq
\pmm{x}{y}{z}.
\eeq
 All isometries preserve the orientation of $\Nil$.

\subsubsect{$\H2\times\R$}
\label{eight:h2r}
 The standard metric is
\beq
ds^2={1\over y^2}(dx^2+dy^2)+dz^2
\eeq
where
$x,y,z\in\R$ and $y>0$.
 The isometry group is $\Isom\H2\times\Isom\R$ hence
an isometry is written as $(\alpha,\beta)$ where $\alpha\in\Isom\H2,
\beta\in\Isom\R$.

\subsubsect{$\tiSLR2$}
\label{eight:sl2r}
 $\tiSLR2$ is the universal covering group of $\SLR2$.
 The standard metric on $\tiSLR2$ is one of left invariant metrics.

\medskip
 This Riemannian manifold can be regarded as the universal cover
of the unit tangent bundle $U\H2$ of the hyperbolic 2-surface $\H2$.
 A tangent bundle of a Riemannian manifold has a natural metric
induced
from the base manifold.
 The induced metric is completely
determined by the metric on the base space.
 So an isometry on the base space induces an isometry on the tangent
bundle.
 We do not explain how to construct the metric on the tangent bundle
in
general and simply show the explicit form in the case of $\H2$.

\medskip
 Let us parametrize $U\H2$ by $(x,y,z)$ where $(x,y)\in\H2$ and
$0\leq z<2\pi$.
 In terms of the tangent bundle $T\H2$, the point $(x,y,z)$
in $\wti{U\H2}$ denotes the tangent vector
$(X,Y)=(y\cos z,y\sin z)$ at $(x,y)\in\H2$.
 Let us denote $(x,y)=(x^1,x^2)$, $(X,Y)=(X^1,X^2)$ and
the metric on $\H2$ by
${}^{(2)}h_{\mu\nu}$.
 We define the covariant total derivative of a tangent vector
$(X^1,X^2)$ by
\beq
DX^{\mu}=dx^{\nu} {\ }^{(2)}\nabla_{\nu} X^{\mu} +dz\ \pd{X^{\mu}} z
\eeq
where
${}^{(2)}\nabla_{\nu}$ is the covariant derivative
associated with ${}^{(2)}h_{\mu\nu}$ and
Greek indices run from 1 to 2.
 The metric on $U\H2$ induced from $\H2$ is
\beqa
\label{eq:sl2r}
ds^2&=&{\ }^{(2)}h_{\mu\nu}dx^{\mu}dx^{\nu}
      +{\ }^{(2)}h_{\mu\nu}DX^{\mu}DY^{\nu}\nono\\
    &=&{1\over y^2}(dx^2+dy^2+(dx+ydz)^2).
\eeqa
 We get the universal cover $\wti{U\H2}$ by simply
replacing the condition $0\leq z<2\pi$ by the condition $z\in\R$.
 The isometry group of \tiSLR2 is 4-dimensional which includes
\tiSLR2.
\tiSLR2 is induced from the isometry group P\SLR2 of the base.
 The additional isometries come from the maps which rotate the unit
tangent
vectors by the same angle,
i.e. send $(x,y,z)$ to $(x,y,z+c)$ with some $c\in\R$.

\subsubsect{$\H3$}
\label{eight:h3}
 $\H3$ is the 3-dimensional hyperbolic space which has the standard
metric
\beq
ds^2={1\over z^2} (dx^2+dy^2+dz^2).
\eeq

 An isometry is well expressed in terms of quaternion.
 A quaternion $q$ is a number which can be written as
$q=a+b i+c j+d k (a, b, c, d \in \R)$.
 All quaternions form a non-Abelian field.
 The multiplication rules for $i, j$, and $k$ are
$i j=-j i=k, j k=-k j=i, k i= -i k=j$.
 An isometry of $\H3$ is expressed as
\beq
q \mapsto (a q+b)(c q+d)^{-1}
\eeq
where
\beq
q=x+y i+z j, z>0
\eeq
and
\beq
a, b, c, d \in \C, a d-b c=1.
\eeq
 These transformations form a Lie group $\PSLC2=\SLC2/\{\pm \bf 1\}$.
 All isometries preserve the orientation of $\H3$.

\subsubsect{\Sol}
\label{eight:sol}
 $\Sol$ is the 3-dimensional group with the following multiplication
rule:
\beq
\left(\ba{c}a\\b\\c\ea\right)\left(\ba{c}x\\y\\z\ea\right)
=\left(\ba{l}a+e^{-c}x\\b+e^c y\\c+z\ea\right).
\eeq
 A basis of left invariant 1-forms are
\beq
\sig^1=e^z dx, \sig^2=e^{-z} dy, \sig^3=dz.
\eeq
 The standard metric is one of left-invariant metrics and is given by
\beqa
ds^2&=&(\sig^1)^2+(\sig^2)^2+(\sig^3)^2  \nono\\
    &=&e^{2z}dx^2+e^{-2z}dy^2+dz^2.
\eeqa
 The identity component of the isometry group is $\Sol$ itself.
 The discrete isometries are
\beqa
& &\pmpmp{x}{y}{z},\\
& &\rpmpmm{x}{y}{z}.
\eeqa
 Therefore $\Isom\Sol$ has eight components.
 Four of them connected to the following elements are orientation
preserving:
\beqa
& &(x,y,z)\mapsto(x,y,z),\\
& &\mmp{x}{y}{z},\\
& &\rppm{x}{y}{z},\\
& &\rmmm{x}{y}{z}.
\eeqa

\subsubsect{$\S3$}
\label{eight:s3}
 $\S3$ is the unit 3-sphere, the unit sphere embedded in $\E4$,
and has a metric induced from $\E4$.
 So the isometry group is \O4 and its orientation preserving subgroup
is \SO4.
 $\S3$ is diffeomorphic to \SU2.
 It is known that \SO4 is isomorphic to $(\SU2\times\SU2)/\ZZ2$,
i.e. there is a homomorphism $\Phi$ from $\SU2\times\SU2$ onto $\SO4$
and the kernel is $\{\pm\bf 1\}$.
 The action of $(q_1,q_2)\in\SU2\times\SU2$ on $p\in\S3$ is
$p\mapsto q_1pq_2^{-1}$ where multiplications are those of $\SU2$.
 Indeed the action of $(q_1,q_2)$ and $(-q_1,-q_2)$ on $\SU2$ are the
same.

\medskip
 Let us introduce a coordinate system.
 An element of $\SU2$ is written as
\beq
p=e^{\alpha \chi_3/2}e^{\beta \chi_2/2}e^{\gamma \chi_3/2}
\eeq
where
\beq
\chi_1=\left(\ba{cc}0&-i\\-i&0\ea\right),\;
\chi_2=\left(\ba{cc}0&1\\-1&0\ea\right),\;
\chi_3=\left(\ba{cc}-i&0\\0&i\ea\right),
\eeq
and
\beq
0\le\alpha<4\pi,\; 0\le\beta<2\pi,\; 0\le\gamma<\pi.
\eeq
 A basis of left-invariant 1-forms $\{\sigma^i\}$ is defined as the
component of the matrix $p^{-1}dp$ of left-invariant 1-forms:
\beq
p^{-1}dp=\sigma^i{\chi_i\over2}.
\eeq
 The basis $\{\sigma_i\}$ of left-invariant 1-forms is
\beqa
& &\sigma_1=-\sin\beta\cos\gamma d\alpha+\sin\gamma d\beta,\nono\\
& &\sigma_2=\sin\beta\sin\gamma d\alpha+\cos\gamma d\beta,\nono\\
& &\sigma_3=\cos\beta d\alpha+d\gamma.
\eeqa

\medskip
 The standard metric is bi-invariant in the point of view that $\S3$
is
identified with \SU2.
 It can be written as
\beqa
ds^2&=&(\sigma^1)^2+(\sigma^2)^2+(\sigma^2)^2\nono\\
    &=&d\alpha^2+d\beta^2+d\gamma^2+2\cos\beta d\alpha d\gamma.
\eeqa

\subsubsect{$\S2\times\R$}
\label{eight:s2r}
 $\S2 \times \R$ is a direct product of a 2-sphere $\S2$ and a line
$\R$.
 $\S2$ can be considered as a unit sphere in $\E3$ which has the
metric
$dl^2$ induced from $\E3$.
 The standard metric of $\S2 \times \R$ is
\beq
ds^2=dl^2+dz^2.
\eeq
 The isometry group of $\S2$ is $\O3$ and its orientation preserving
subgroup
is $\SO3$.
 An isometry of $\R$ is a translation or a reflection at a point.
 A translation preserves the orientation of $\R$ while a reflection
reverses it.
 The isometry group of $\S2 \times \R$ is $\Isom\S2 \times \Isom\R$,
i.e.
an isometry is expressed as $(\alpha,\beta)$ where $\alpha$ is an
element
of \O3 and $\beta$ is a translation or a reflection of $\R$.
 Since both $\Isom\S2$ and $\Isom\R$ have two components,
$\Isom(\S2 \times \R)$ has four components.
 The identity component is $\SO3 \times \R$.
 An isometry $(\alpha,\beta)$ preserves the orientation of
$\S2\times\R$
when both $\alpha$ and $\beta$ are either orientation preserving
or orientation reversing.
 Therefore $\Isom^+ (\S2 \times \R)$ has two components.

\subsect{Homogeneity preserving diffeomorphisms}
\label{hpd}
 Let us consider an especially important class of diffeomorphisms,
homogeneity preserving diffeomorphisms \cite{AS}.

\medskip
 Let $(X,G)$ be a geometry.
 A diffeomorphism $\phi$ on $X$ is a {\it homogeneity preserving
diffeomorphism}\/ (HPD) if it preserves the action of $G$ on $X$,
i.e. it is the self-equivalence map of a geometry $(X,G)$.
 If a diffeomorphism $\phi$ is an HPD then $\bar{\phi}:g \mapsto \phi
\circ g \circ \phi^{-1} $ is an automorphism of $G$, i.e.
\begin{equation}
\bar{\phi} \in {\rm Aut}(G).
\end{equation}
 Group theoretically, an HPD is an element of the normalizer of
G in \mbox{Diff}$X$, the group of all diffeomorphisms.

\medskip
 A diffeomorphism $\phi$ induces a map $\phi_*:TX \to TX$.
 We show that $\phi_*$ gives an automorphism of Lie algebra of $G$.
 Let Exp$t\xi$ denote the 1-parameter group
of transformations generated by
a vector field $\xi$ on $X$.
 Let Alg($G$) denote the Lie algebra of $G$, the space of
 vector fields on $X$ which generate transformations in $G$.
 By definition the condition for $\phi$ to be an HPD is that
$\bar{\phi}:
{\rm Exp} t A \mapsto \phi \circ {\rm Exp}tA \circ \phi^{-1}$ is an
automorphism of $G$, where $A\in\mbox{Alg}(G)$.
 For any vector field $\xi$ on $X$ and a diffeomorphism $\phi$,
\begin{equation}
\phi \circ {\rm Exp}t\xi\circ \phi^{-1}={\rm Exp}t\phi_*\xi
\end{equation}
holds, i.e. the generator of the 1-parameter family of
diffeomorphisms on the
left hand side is $\phi_*\xi$.
 So in particular for any $A \in {\rm Alg}(G)$ we have
\begin{equation}
\phi \circ {\rm Exp}tA \circ \phi^{-1}={\rm Exp}t\phi_*A.
\label{eq:Exp}
\end{equation}
 Therefore $\phi$ is an HPD if and only if
\begin{equation}
\phi_*{\rm Alg}(G)={\rm Alg}(G).
\end{equation}
Since the restriction of $\phi_*$ on ${{\rm Alg}(G)}$ is a linear map
and preserves the
commutator of vector fields, it is a Lie algebra automorphism, i.e.
\begin{equation}
\phi_* \in \mbox{Aut(Alg}(G))
\end{equation}
where Aut(Alg($G$)) is the group of all Lie algebra automorphisms of
Alg($G$).

\medskip
 Note that the converse is not true.
 A Lie algebra automorphism is not always induced from an HPD because
an HPD
has to be globally defined on $X$.

\sect{Compact locally homogeneous models}
\label{def}
 In this section we define the compact locally homogeneous universes
and their physical degrees of freedom.

 A spacetime is represented by a pair $(^{(4)}M,g_{ab})$, where
$^{(4)}M$
is a 4-manifold and $g_{ab}$ is a Lorentzian 4-metric.
 We 3+1 decompose the spacetime, i.e. we think of a
spacetime $(^{(4)}M,g_{ab})$ as the time evolution of a Riemannian
3-metric $h_{ab}(t)$ on a 3-manifold $M$.
 We require that this decomposition is synchronous;
the lapse function is unity and the shift vector vanishes.
Then the 4-metric can be written as
\begin{equation}
g_{ab}=-\nabla_a t\nabla_b t + h_{ab}.
\end{equation}
 We say a spacetime $(M,h_{ab}(t))$ to be  {\it spatially locally
homogeneous}\/ if $h_{ab}(t)$ for every $t$ is locally homogeneous.
 We say $(M,h_{ab}(t))$ to be {\it spatially compact}\/ if $M$ is
compact.

 Let $(M,h_{ab}(t))$ be a spatially compact locally homogeneous
spacetime.
 Let us separate the conformal factor $a(t)$ from $h_{ab}(t)$ as
\begin{equation}
h_{ab}(t)=a^2(t) \bar{h}_{ab} (t)
\end{equation}
 and think of $\bar{h}_{ab}$ as the dynamical variables.
 Therefore, the physical degrees of freedom are the numbers
needed to parametrize $\bar{h}_{ab}$.
 Hereafter we omit the bar of $\bar{h}_{ab}$ and simply denote it
by $h_{ab}$.

\medskip
 As mentioned in the previous section, $M$ has a unique
universal cover $\ti{M}$ up to diffeomorphisms,
and $(M,h_{ab}(t))$ is metrically diffeomorphic to
$(\ti{M},\ti{h}_{ab}(t))/\Gamma (t)$
where $\ti{h}_{ab}$ is homogeneous and $\Gamma$ is a discrete
subgroup of Isom$(\ti{M},\ti{h}_{ab})$.
 All $\Gamma (t)$ must be isomorphic for all $\ti{M}/\Gamma (t)$
to have the same topology.

\medskip
 Thinking of a manifold $M$ as a quotient $\ti{M}/\Gam$ of a simply
connected homogeneous manifold $\ti{M}$ naturally leads us to
separating
the degrees of freedom of $h_{ab}$ into
those of $\ti{h}_{ab}$ and those of $\Gamma$, i.e.
those of the universal cover $(\ti{M},\ti{h}_{ab})$ and the
Teichm\"{u}ller deformations.
 Here we should recall that there is freedom of diffeomorphisms
in giving a universal cover $\ti{M}$ of $M$.
 If $\phi$ is an orientation preserving  diffeomorphism
from $\ti{M}$ onto itself,
$(\ti{M},\ti{h}_{ab})/\Gamma$ and
$(\ti{M},\phi_* \ti{h}_{ab})/\phi \circ \Gamma \circ \phi^{-1}$
are diffeomorphic metrically and have the same orientation.
 This means that different pairs of $\ti{h}_{ab}$ and $\Gamma$
may give the same manifold $(M,h_{ab})$.
 To avoid this uninteresting freedom emerged from diffeomorphisms,
and recalling that we have already put aside the conformal factor,
we define the degrees of freedom of the universal covering manifold
as those of
\beq
\label{eq:univdof}
{\mbox{\{homogeneous metrics on a simply connected manifold which
admit compact quotients\}}
\over
\mbox{\{orientation preserving diffeomorphisms\}
\{global conformal transformations\}}}.
\eeq
 We say that there are degrees of freedom of the universal cover
if these equivalence classes can be smoothly deformed.

\medskip
 Once a metric on the universal cover $\ti{M}$ is fixed, the isometry
group of $\ti{M}$ is determined.
 Remaining deformations of the manifold $M$ are the \Teich
deformations.

\medskip
 We investigate all homogeneous universal covers in Sec.\ref{univ}
and all compact quotients and their \Teich spaces in
Sec.\ref{quotients}.

\sect{Types of universal covers}
\label{univ}
 In this section we give all equivalence classes of homogeneous
metrics mentioned at the end of Sec.\ref{def}.
 We explain the method of exhausting all of them
in Sec.\ref{univ:method}, and
we give them explicitly in Sec.\ref{univ:results}.

\subsect{The method for obtaining nonmaximal universal covers}
\label{univ:method}
 Theorem \ref{th:thurston} tells us all simply connected
{\it maximal} geometries which admits compact quotients.
 But we want to consider a general locally homogeneous metric on
a compact manifold,
so the requirement that the geometry be maximal is too strong for our
purpose. For example, Bianchi IX space does not satisfy this
requirement;
the geometry ($S^3$(topologically), SU(2)) is a subgeometry of
the isotropic 3-sphere ($S^3$, SO(4)).
 In general, if $(X,G\pr)$ is a subgeometry of $(X,G)$ then
$G\pr$ is a subgroup of $G$ and acts transitively on $X$.
 It follows that the the stabilizer of $G\pr$ is a subgroup
of that of $G$.
 In other words, a metric on a maximal geometry is more isotropic
than those on any of its subgeometries.

\medskip
 Theorem \ref{th:thurston} says nothing about nonmaximal geometries
which have
compact quotients.
 Nevertheless we can show that we have only to consider subgeometries
of the
above eight geometries.
 Let $(X,G\pr)$ be a simply connected 3-geometry which admits
a compact quotient.
Then there is a discrete freely acting subgroup
$\Gamma\pr$ of $G\pr$ which makes $X/\Gam\pr$ compact.
 Let $(X,G)$ be a maximal geometry which has $(X,G\pr)$
as its subgeometry.
 Then $\Gam\pr$ is a subgroup of $G$ which makes $X/\Gam\pr$
 compact, so our maximal, simply connected geometry $(X,G)$
admits compact quotients.
 By Theorem \ref{th:thurston} our maximal geometry $(X,G)$ is
one of the eight Thurston geometries, which proves that
$(X,G\pr)$ is a subgeometry of one of the eight Thurston geometries.

\medskip
 Let us go back to Riemannian manifolds.
 If a Riemannian manifold $(\ti{M},\ti{h}_{ab})$ admits compact
quotients then the geometry $(\ti{M},\Isom^+\ti{M})$ is a subgeometry
of
the eight Thurston maximal geometries.
 Our strategy is to consider all subgeometries $(\ti{M},G)$ of the
Thurston
geometries and $G$-invariant metrics on $\ti{M}$.
 Here we must recall that $G$ may not be the isometry group itself
, i.e. it may be the
case that the $G$-invariant metric on $\ti{M}$ admits a larger
isometry group than $G$.
 So it is necessary to find the whole $\Isom^+\ti{M}$ in each case.
 Of course, if $(M,\ti{h}_{ab})$ and $(M,\ti{h}\pr_{ab})$ are in the
same equivalence class then $\Isom^+(\ti{M},\ti{h}_{ab})$
and $\Isom^+(\ti{M},\ti{h}\pr_{ab})$ are isomorphic.

\medskip
 We adopt the following procedure to find all equivalence classes of
homogeneous metrics on a simply connected manifold which admit
compact
quotients and their isometry groups:

 1)List up all subgeometries $(\ti{M},G)$ of each of the Thurston
geometries.

 2)For each subgeometry $(\ti{M},G)$, form a general $G$-invariant
Riemannian metric
$\ti{h}_{ab}^\prime$ on $\ti{M}$.

 3)Transform the metric into a certain simple `representative metric'
$\ti{h}_{ab}$ by diffeomorphisms together with global conformal
transformations and find the equivalence classes.

 4)Find the isometry group of $(\ti{M},\ti{h}_{ab})$.

\noindent
 We thus get all possible universal covers of locally homogeneous
Riemannian manifolds.
 Logically speaking, equivalence classes thus found
do not necessarily have to admit compact quotients.
 So we have another task.

5)Make sure whether the equivalence class really admits compact
quotients.

 The Bianchi classification \cite{LL,Wa} is of great use
in carrying out the procedure
given above,
for it is the classification of all simply connected 3-dimensional
Lie groups up to isomorphisms.
A correspondence of the Bianchi classification to the eight Thurston
geometries
was first pointed out by Fagundes~\cite{Fa}.
We elaborate this correspondence precisely in terms of geometry
in order to make it applicable for our program.

\medskip
 Let us say a geometry $(X,G)$ to be {\it minimal}\/ if it does not
have a proper subgeometry.
 We can say that the Bianchi classification is the classification
of 3-dimensional minimal geometries,
for the following fact is well known:
\begin{fact}
Any minimal, simply connected 3-dimensional geometry
is equivalent to $(X,G)$
where $X=\R^3$, $G=$one of Bianchi I to Bianchi VIII groups;
$X=\S3$, $G=$Bianchi IX group;
or $X=\S2\times\R$, $G=\SO3\times\R$.
\end{fact}
 Let us call the minimal geometries with Bianchi groups the {\it
Bianchi
minimal geometries}, the last one the {\it Kantowski-Sachs-Nariai
{\rm (}KSN{\rm )} minimal geometry}, and in all the {\it Bianchi-
Kantowski-Sachs-Nariai {\rm (}BKSN{\rm )} minimal geometries}.

 We emphasize that
the Thurston geometries are the maximal 3-dimensional geometries
which give compact quotients
while the BKSN geometries are the minimal 3-dimensional geometries.

\medskip
 Let us categorize Thurston's geometries $(X,\Isom X)$ by the
topology of $X$:
the one with $X\cong\R^3$ i.e. $X=\E3, \H3, \H2\times\R, \tiSLR2,
\Nil$, or $\Sol$;
the one with $X=\S3$; and the one with $X=\S2\times\R$.
 A Thurston geometry in the first case has at least one of Bianchi I
to
Bianchi VIII geometries as its subgeometry.
 The geometry $(\S3,\Isom^+\S3)$ has Bianchi IX
geometry as its subgeometry.
 The geometry $(\S2\times\R,\Isom(\S2\times\R))$ has the KSN
geometry as its subgeometry.

\medskip
 We explain the procedure written above in the case of the Bianchi
minimal geometries.
 The case of the KSN geometry is rather easy and is
discussed separately in Sec.\ref{univ:results}.
 There are two convenient feature when $(X,G)$ is a Bianchi geometry.
 First, we can deal with the $G$-invariant 1-forms $(\sigma\pr{}^i)_a
(i=1,2,3)$.
 Any $G$-invariant metric $\ti{h}\pr_{ab}$ can be written as
\begin{equation}
\label{eq:h=hsigsig}
\ti{h}\pr_{ab}=\ti{h}\pr_{ij} (\sigma^i)_a (\sigma^j)_b
\end{equation}
with $\ti{h}\pr_{ij}$ being a constant matrix.
 Second, a Lie algebra automorphism is always induced from a
diffeomorphism
which is globally defined.

\medskip
 We discuss homogeneity preserving diffeomorphisms (HPDs),
because they play an essential role in reducing
the redundant degrees of freedom of the metric.
 We have already seen  in Sec.\ref{hpd} that an HPD $\phi$
on a geometry $(\ti{M},G)$
induces a Lie algebra automorphism of Alg$(G)$, by which
the invariant 1-forms transform as
\beq
\label{eq:f}
(\sigma\pr{}^i)_a=f^i{}_j(\sigma^j)_a \mapsto (\sigma^i)_a
\end{equation}
where $f^i{}_j$ is a constant $3\times3$ matrix
and $\sigma^i$ and $\sigma\pr{}^i$ must satisfy the Maurer-Cartan
equations,
\begin{eqnarray}
\label{eq:str}
d\sigma^i&=&-{1 \over 2}C^i{}_{jk}\sigma^j \wedge \sigma^k,        \\
\label{eq:str'}
d\sigma\pr{}^i&=&-{1 \over 2}C^i{}_{jk}\sigma\pr{}^j \wedge
\sigma\pr{}^k.
\end{eqnarray}
 Substituting (\ref{eq:f}) into (\ref{eq:str'}) and using
(\ref{eq:str}), we have
\begin{equation}
\label{eq:fc=cff}
f^i{}_j C^j{}_{lm}=C^i{}_{jk} f^j{}_l f^k{}_m.
\end{equation}

\medskip
 Conversely, given a Lie algebra automorphism $f^i{}_j$ which satisfy
(\ref{eq:fc=cff}) the corresponding HPD is obtained by solving a
differential equation.
 We must introduce a coordinate system $x=\{x^i\}\ (i=1,2,3)$ to
denote
a point $p$ in $\ti{M}$.
 Let $\phi^i(x)$ denote the $i$-th coordinate of $\phi(p)$.
 Then we can expand $\sigma^i$ by $dx^i$ as
\beq
\sigma^i=\sigma^i{}_j(x) dx^j.
\eeq
 The primed one is the right hand side with $\{x^i\}$ replaced by
$\{\phi^i\}$:
\beqa
\sigma\pr{}^i&=&\sigma^i{}_j(\phi(x)) d\phi^j      \nono\\
             &=&\sigma^i{}_j(\phi(x)) \Del {\phi^j}{x^k} dx^k.
\eeqa
 It follows from $\sigma\pr{}^i=f^i{}_j \sigma{}^j$ that
\beq
\sigma^i{}_j(\phi(x)) \Del {\phi^j}{x^k}=f^i{}_j \sigma^j{}_k(x).
\eeq
 So the Jacobi matrix is
\beq
\label{eq:jacobi}
\Del{\phi^i}{x^j}=(\sigma^{-1})^i{}_k(\phi(x))\ f^k{}_l\
\sigma^l{}_j(x).
\eeq
The HPD is the solution to this differential equation.

\medskip
 Analytically, eq.(\ref{eq:fc=cff}) is the local integrability
condition for
eq.(\ref{eq:jacobi}).
 If $G$ is a Bianchi group an HPD $\phi$ can be identified with
a Lie group automorphism $\bar\phi$
since the manifold $\ti{M}$ is diffeomorphic to $G$.
 Moreover, there is a correspondence between Lie group automorphisms
$\bar\phi$
and Lie algebra automorphisms $f^i{}_j$, for there is a one-to-one
correspondence
between simply connected Lie groups and Lie algebras.
 Therefore, in the case of the Bianchi geometries, the necessary
and sufficient condition for $f^i{}_j$
to be induced from an HPD on $\ti{M}$
is that eq.(\ref{eq:fc=cff}) holds.

\medskip
 Let us reduce the redundant degrees of freedom of
a general $G$-invariant metric
by the degrees of freedom of HPDs and global conformal
transformations.
 An HPD $\phi$ transforms the invariant 1-forms
$(\sigma\pr{}^i)_a$ as (\ref{eq:f}).
 By (\ref{eq:h=hsigsig}) the metric transforms as
\begin{eqnarray}
\label{eq:h'->h}
\ti{h}\pr_{kl} (\sigma^i)_a(\sigma^j)_b
&=&f^k{}_i f^l{}_j \ti{h}_{kl} (\sigma^i)_a(\sigma^j)_b  \nono\\
&=&\ti{h}_{ij}(\sigma\pr{}^i)_a(\sigma\pr{}^j)_b         \nono\\
&\mapsto&\ti{h}_{ij}(\sigma^i)_a(\sigma^j)_b.
\end{eqnarray}
 For any metric $\ti{h}\pr_{ij}$ we find a transformation
$\ti{h}\pr_{ij}$
as (\ref{eq:h'->h}) by diffeomorphisms
to get as simple a metric $\ti{h}_{ij}$ as possible.

\medskip
 It is more convenient to deal not with $\ti{h}_{ij}$ itself but
with triad 1-form $(e^i)_a$.
 A triad 1-form of the metric $\ti{h}\pr_{ab}$ is
\begin{equation}
\label{eq:triad}
(e^i)_a=b\pr{}^i{}_j(\sigma^i)_a
\end{equation}
where
\begin{equation}
\label{eq:h=bb}
\ti{h}\pr_{ij}=\delta_{kl} b\pr{}^k{}_i b\pr{}^l{}_j.
\end{equation}
 We are going to reduce the degrees of freedom of $b\pr{}^i{}_j$
by using the degrees of freedom of $f^i{}_j$ together with a global
conformal
transformation $b\pr{}^i{}_j=a b^i{}_j \mapsto b^i{}_j$ to obtain a
simple $b^i{}_j$.
 Note that infinitely many $b\pr{}^i{}_j$ correspond to the same
$\ti{h}\pr_{ij}$; in fact $b\pr{}^i{}_j$ and $r^i{}_k b\pr{}^k{}_j$
correspond to the same $\ti{h}\pr_{ij}=\delta_{kl} b\pr{}^k{}_i
b\pr{}^l{}_j$
if $r^i{}_j \in \O3$.
 Therefore the condition for
$\ti{h}\pr_{ij}=\delta_{kl} b\pr{}^k{}_i b\pr{}^l{}_j$
to be any positive definite symmetric matrix is
for $b\pr{}^i{}_j$ to be any element of $\O3\backslash\GLR3$.
 The transformation law for $b\pr{}^i{}_j$ is
\begin{equation}
\label{eq:b'->b}
 b\pr{}^i{}_j=b^i{}_k f^k{}_j \mapsto b^i{}_j.
\end{equation}
 We must find the matrix $b^i{}_j$ such that
\begin{equation}
\label{eq:b'=afb}
b\pr{}^i{}_j=a b^i{}_k f^k{}_j
\end{equation}
expresses any element of $\O3\backslash\GLR3$, i.e.
\begin{equation}
\label{eq:arbf}
r^i{}_k b\pr{}^k{}_j=a r^i{}_k b^k{}_l f^l{}_j
\end{equation}
expresses any element of $\GLR3$, where $a>0, r^i{}_j \in \O3$
and $f^i{}_j$ satisfies (\ref{eq:fc=cff}).

\medskip
 The number $U$ of degrees of freedom of the universal cover is
\bea
U&=&\left(\ba{l}\mbox{degrees of}\\ \mbox{freedom of}\\
  \GLR3\ea\right)
-\left(\ba{l}\mbox{degrees of}\\ \mbox{freedom of}\\r^i{}_k
  \ea\right)\nono\\
& &-\left(\ba{l}\mbox{degrees of}\\ \mbox{freedom of global}\\
\mbox{conformal transformations} \ea\right)
-\left(\ba{l}\mbox{degrees of}\\ \mbox{freedom of}\\
\mbox{Lie algebra automorphisms} \ea\right) \nono\\
&=&9-3-1-\mbox{dim(Aut(Alg$G$))}\nono\\
\label{eq:dof:univ}
&=&5-\mbox{dim(Aut(Alg$G$))}
\eea
for the Bianchi geometries except for Bianchi I geometry.
 A Bianchi I group-invariant metric has global conformal
transformations.
 The degrees of freedom of conformal transformations (=1) is included
in
those of diffeomorphisms so that
\beq
U=6-\mbox{dim(Aut(Alg$G$))}.
\eeq
 It follows from dim(Aut(Alg$G$))=9 that $U=0$.

\subsect{Types of universal covers}
\label{univ:results}
 Now we show the procedure shown in Sec.\ref{univ:method} in detail
for a few of the Thurston maximal geometries
and show the results for the others.
 In this section
we find the equivalence classes
which are slightly different from (\ref{eq:univdof}), where
`orientation preserving diffeomorphisms' is replaced by
`diffeomorphisms'.
 In Sec.\ref{univ:summary} we find the equivalence classes
(\ref{eq:univdof}) as a corollary.

\subsubsect{$(\E3,\Isom\E3)$}
\label{univ:e3}
 We have seen in Sec.\ref{eight} that $\Isom\E3=\IO3$.
A subgroup of $\IO3$ which acts transitively on $\R^3$ must contain
the
translations in (\ref{eq:iso3}), i.e. the corresponding Lie algebra
must
contain $K_1,K_2$, and $K_3$.
The basis of possible Lie subalgebras up to isomorphisms are the
following:
\begin{eqnarray}
1)& & A_1=K_1, A_2=K_2, A_3=K_3;  \\
\label{eq:alg:vii0}
2)& & A_1=K_1, A_2=K_2, A_3=K_3+K_6;  \\
3)& & A_1=K_1, A_2=K_2, A_3=K_3, A_4=K_6;\\
4)& & A_i=K_i \ \ \ (1\leq i \leq6.)
\end{eqnarray}
The connected Lie groups corresponding to above Lie algebras are
1)$\R^3$ (all translations) i.e. Bianchi I group,
2)Bianchi VII(0) group,
3)$\SO2 \times \R^3$, and
4)$\ISO3$.
The groups corresponding to minimal geometries are $\R^3$ and Bianchi
VII(0)
group.
 $\SO2 \times \R^3$ contains $\R^3$ as its subgroup and $\ISO3$
contains
both $\R^3$ and Bianchi VII(0) group.
 Let us consider each of them in detail.

\medskip
1)$\R^3$.

The invariant 1-forms of $\R^3$ are
\beq
\label{eq:r3:form}
\sig^i=d x^i\ \ (i=1,2,3)
\eeq
where $x^1=x, x^2=y, x^3=z$.
Any $\R^3$-invariant metric can be written as
\beq
\ti{h}_{ab}=\ti{h}_{ij}(\sig^i)_a(\sig^j)_b.
\eeq
Since all structure constants of $\R^3$ vanish, (\ref{eq:fc=cff}) is
trivial.
 This means that $f^i{}_j$ can be any element of $\GLR3$.
In fact, by (\ref{eq:f}) and (\ref{eq:r3:form}) the matrix $f^i{}_j$
is
the Jacobi matrix itself of an HPD so that any HPD is given by
the solution of (\ref{eq:jacobi}):
\beq
x^i\mapsto f^i{}_j x^j+a^i,
\eeq
where $f^i{}_j\in \GLR3$ and $a^i$ is a constant vector.
 We can make $b^i{}_j$ to be an identity matrix, which is enough for
(\ref{eq:arbf}) to express a general element of $\GLR3$.
 Any $\R^3$-invariant metric is in the same equivalence
class as the representative metric
\beq
\label{eq:r3:rep}
d s^2=dx^2+d y^2+d z^2.
\eeq
The isometry group of this metric is $\IO3$, as given in
Sec.\ref{eight}.

\medskip
 Note especially that any $\R^3$-invariant metric is
$\IO3$-invariant, i.e.
its isometry group is isomorphic to $\IO3$.
 In this case HPDs are translations, rotations, and stretches.
 Translations of course do not change the metric because they are
isometries.
 Using degrees of freedom of other two types of HPDs we can transform
the
metric to the standard metric.
 In fact, there is no need of global conformal transformations in
this
case.
 This comes from the fact that $\E3$ has global conformal isometries.
 ($\E3$ is the only such a manifold.)

\medskip
 2)Bianchi VII(0) group.

The invariant 1-forms of Bianchi VII(0) group are
\beq
\label{eq:vii0:form}
\sig^1=\cos z dx+\sin z dy,\ \sig^2=-\sin z dx+\cos z dy,\ \sig^3=dz.
\eeq
 The nonvanishing structure constants are
\beq
C^1{}_{23}=-C^1{}_{32}=C^2{}_{31}=-C^2{}_{13}=1.
\eeq
Solving (\ref{eq:fc=cff}) we have either
\beqa
\label{eq:vii0:f..1}
f^2{}_1=-f^1{}_2, f^2{}_2=f^1{}_1, f^3{}_1=f^3{}_2=0, f^3{}_3=1
\eeqa
or
\beqa
\label{eq:vii0:f..2}
f^2{}_1=f^1{}_2, f^2{}_2=-f^1{}_1, f^3{}_1=f^3{}_2=0, f^3{}_3=-1.
\eeqa
 The simplest choice of the components of $b^i{}_j$
for (\ref{eq:arbf}) to express a general element of $\GLR3$
is
\beqa
\label{eq:vii0:b}
& &b^1{}_1=\alpha, b^2{}_2=\alpha^{-1}, b^3{}_3=1,\nono \\
& &b^i{}_j=0 \mbox{   (the others),}
\eeqa
where $\alpha\ge 1$.
 If we rewrite the parameter $\alpha$ as $\alpha=e^\lam$
the representative metric is
\beq
\label{eq:vii0:rep}
ds^2= \aaa (\cos z dx +\sin z dy)^2 +
\aam (-\sin z dx +\cos z dy)^2+dz^2
\eeq
where $\lambda\ge 0$.
 The equivalence classes form a 1-parameter family.
 Note that the metric (\ref{eq:r3:rep}) is the special case
($\lam=0$) of the metric (\ref{eq:vii0:rep}).

The isometry is either
\beqa
\label{eq:vii0:isom+}
\phi^1(x)&=&\pm (x\cos c^3-y\sin c^3)+c^1,\nono\\
\phi^2(x)&=&\pm (x\sin c^3+y\cos c^3)+c^2,\nono\\
\phi^3(x)&=&z+c^3
\eeqa
or
\beqa
\label{eq:vii0:isom-}
\phi^1(x)&=&\pm (x\cos c^3-y\sin c^3)+c^1,\nono\\
\phi^2(x)&=&\mp (x\sin c^3+y\cos c^3)+c^2,\nono\\
\phi^3(x)&=&-z+c^3
\eeqa
where $c^i (i=1,2,3)$ are constants.
 The isometry group have four components.
 All isometries are orientation preserving.
 The identity component of the isometry group
is given by taking the positive signs of (\ref{eq:vii0:isom+}).
 It is Bianchi VII(0) group.

\medskip
 3)$\SO2\times\R^3$.

 The group $\SO2 \times \R^3$ has $\R^3$ as its subgroup.
 So from 1) any $\SO2 \times \R^3$-invariant
metric is in the same equivalence class as the metric
$d s^2=dx^2+d y^2+d z^2$, and hence is $\IO3$-invariant.

\medskip
 4)\ISO3.

 Any $\ISO3$-invariant metric is $\IO3$-invariant.

\subsubsect{(\Nil,\Isom\Nil)}
 Nil is the only minimal subgroup of the group \Isom\Nil.
 All Nil-invariant metrics are in the same equivalence
class as the representative metric
\beq
d s^2 = dx^2+dy^2+(dz-xdy)^2.
\eeq
 Its isometry group is already given explicitly in
Sec.\ref{eight:nil}.
 The isometry group is 4-dimensional.

\subsubsect{$(\H2\times\R,\Isom(\H2\times\R))$}
 The subgeometries of the geometry $(\H2\times\R,\Isom(\H2\times\R))$
with connected Lie groups are Bianchi III geometry and
$(\H2\times\R,\Isom_0(\H2\times\R))$.

 The isometry group of a general Bianchi III-invariant metric is
3-dimensional and its identity component is Bianchi III group itself.
 This group does not admit compact quotients (Appendix) and will not
be discussed further.

 Any $\Isom_0(\H2\times\R)$-invariant metric is equivalent to
\beq
ds^2= {1\over{y^2}}(dx^2+dy^2)+dz^2,
\eeq
which is the standard metric appeared in Sec.\ref{eight:h2r}
where its isometry group $\Isom(\H2\times\R)$ is also given.

\subsubsect{$(\tiSLR2,\Isom\tiSLR2)$}
\label{univ:results:slr2}
 Transitively acting, connected subgroups of the group \Isom\tiSLR2
are
Bianchi III group, Bianchi VIII group = \tiSLR2, and
$\Isom_0\tiSLR2$.
 The first two correspond to minimal subgeometries.

 General Bianchi III-invariant metrics do not admit compact
quotients;
general \tiSLR2-invariant metrics do not admit compact quotients
either (Appendix).
 Only $\Isom_0\tiSLR2$-invariant metrics admit compact quotients.

 The representative metric is of the form
\beq
ds^2 = {1\over{y^2}} (\bbb (dx^2+dy^2)+\bbm (dx+y dz)^2),
\eeq
which forms a one parameter family.
 The isometry group is found to be \Isom\tiSLR2 explained in
Sec.\ref{eight:sl2r}.

\subsubsect{$(\H3,\PSLC2)$}
 The group \PSLC2 has two minimal subgroups, Bianchi V group and
Bianchi
VII(A) group.
 Any subgroup of \PSLC2 has either of them as its subgroup.

 The representative metric of the Bianchi V-invariant metrics is
\beq
\label{eq:v:metric}
ds^2 = {1\over z^2} (dx^2+dy^2+dz^2).
\eeq
 Any Bianchi V-invariant metric is \PSLC2-invariant.

 The representative metric of the Bianchi VII(A)-invariant metric is
\beq
\label{eq:viia:metric}
ds^2 = \alpha^2 e^{2Az}(\cos z dx+\sin z dy)^2
      +\alpha^{-2} e^{2Az}(-\sin z dx+\cos z dy)^2
      +dz^2
\eeq
where $\alpha\ge 1$.
 Its isometry group is 3-dimensional and has two components.
 All isometries preserve the orientation.
 The identity component is Bianchi VII(A) group itself.
 The metric (\ref{eq:v:metric}) is a special case of the metric
(\ref{eq:viia:metric}).
 However, Bianchi VII(A) geometry does not admit compact quotients
(Appendix).

\subsubsect{(\Sol,\Isom\Sol)}
 The geometries $(\Sol,\Isom_0\Sol)=(\R^3,\Sol)$
are the only subgeometry of (\Sol,\Isom\Sol) with a connected group.
 The representative metric is
\beq
ds^2 ={1\over2}\aaa(e^z dx+e^{-z}dy)^2+{1\over2}\aam(-e^z
dx+e^{-z}dy)^2+dz^2.
\eeq
 The equivalence classes form a one parameter family.
 The isometry group is $\Isom^+\Sol$ for $\lambda\neq0$ and
$\Isom\Sol$ for $\lambda=0$, which are mentioned in
Sec.\ref{eight:sol}.

\subsubsect{$(\S3,\O4)$}
 The group \O4 has Bianchi IX group i.e. \SU2 as its minimal
subgeometry.
The representative metric is
\beq
ds^2 = \bb (\sigma^1)^2+\cc (\sigma^2)^2+\dd (\sigma^3)^2
\eeq
where $\{\sigma^i\}$ (i=1,2,3) is a basis of left-invariant 1-forms
given in Sec.\ref{eight:s3}.
 The equivalence classes form a two parameter family.
 The isometry group is found to have four components;
the identity component is \SU2 itself.
 The discrete isometries are written as
\beq
p\mapsto p\chi_i\quad (i=1,2,3).
\eeq

\medskip
 If we put $\lambda_-=0$, then the isometry group becomes larger.
 Its identity component is $(\SU2 \times \U1)/\ZZ2$.
 It has two components and the discrete isometry is
\beq
p\mapsto p\chi_3.
\eeq
 Here the action of \U1 is the right multiplication of a matrix
of the form
$\left(\ba{ll}
e^{i\theta} & 0\\
0           & e^{-i\theta}
\ea\right)$
to the element of \SU2.
 If we further put $\lambda_+=0$, then the isometry group is \O4.

\subsubsect{$(\S2\times\R, \Isom(\S2\times\R))$}
\label{univ:s2r}
 The Thurston geometry $(\S2\times\R, \Isom(\S2\times\R))$ has the
KSN
geometry $(\S2\times\R, \SO3\times\R)$ as its minimal subgeometry.
 The group $\SO3\times\R$ is the identity component of
$\Isom(\S2\times\R$).

\medskip
 Let us parametrize points in $\S2\times\R$ by $(r, \theta, z)$
where $0\leq r<\pi, -\pi\leq\theta<\pi$ and $z\in\R$.
 Any $\SO3\times\R$-invariant metric can be written as
\beq
ds^2 =a^2 dl^2 + b^2 dz^2,
\eeq
where $dl^2$ is the metric on $\S2$:
\beq
dl^2=dr^2+\sin^2 r d\theta^2.
\eeq
 Transforming the metric by the diffeomorphism
$(r,\theta,az)\mapsto(r,\theta,bz)$
and by the global conformal transformation
$\ti{h}\pr_{ij}\mapsto \ti{h}_{ij}=a^{-2} \ti{h}\pr_{ij}$,
we get the representative metric
\beq
ds^2 = dl^2 + dz^2.
\eeq
 This is the same as the standard metric in Sec.\ref{eight:s2r}
and hence the isometry group is $\Isom(\S2\times\R$).

\subsect{Summary of the section}
\label{univ:summary}
 Let us summarize the result in this section as a theorem.
\bth
\label{th:univ1}
 Any simply connected 3-dimensional Riemannian manifold
which admits a compact quotient is one of the types in
Table 1 up to diffeomorphism and global
conformal transformation.
\eth

\medskip
 Now we give equivalence classes (\ref{eq:univdof}) which give the
universal
covers of all orientable compact locally homogeneous Riemannian
manifolds.
\bth
\label{th:univ2}
 Any simply connected 3-dimensional Riemannian manifold
which admits a compact quotient is one of the types in
Table 2 up to orientation preserving diffeomorphism and global
conformal transformation.
\eth
 Let us give a brief proof.
 If a manifold $\ti M$ has an orientation reversing isometry
the result is the same as Theorem \ref{th:univ1} above;
if not, $(\ti M,\ti g_{ab})$ and $(\ti M,r_*\ti g_{ab})$
are not in the same equivalence class but have isomorphic isometry
groups,
where $r$ is an orientation reversing diffeomorphism.
 An orientation reversing diffeomorphism is well expressed by the
group
multiplication of the minimal subgeometry for types a to g,
namely $r:p\mapsto p^{-1}$.
 This map carries a left multiplication to a right multiplication.
 It can be easily verified by exhaustion that
there is an orientation reversing isometry if and only if
$r$ is an isometry, i.e. the left-invariant metric is also
right-invariant.
 Types a1, b, f, g1 and g2 are in the case of $\Isom\ti M=\Isom^+\ti
M$.
 In the case of Types a and f, the range of the parameter $\lambda$
must be expanded to $\R$.
 In the case of Types b and g1 the metric
$(\ti g_R)_{ab}\equiv r_*(\ti g_L)_{ab}$
has right invariant with respect to $\Nil$ and \SU2 respectively.
 In the case of Type g2 the metric $(\ti g_R)_{ab}\equiv r_*(\ti
g_L)_{ab}$
is invariant under right action of \SU2 and left action of \U1.

\medskip
 The number of the degrees of freedom is that of parameters
in the representative metric,
which we call {\em universal cover parameters}.
 They
 are also characterized by free parameters appearing in
the relative ratios of the nonvanishing principal sectional
curvatures.
 These ratios can be chosen to be dynamical variables.

\sect{Compact quotients and their \Teich spaces}
\label{quotients}

 We shall give compact quotients modeled on each of the universal
covers from Type a to Type h given in the previous section
and give the dimension $T$ of their \Teich spaces.
 We make use of the knowledge of compact quotients modeled on the
Thurston
maximal geometries and the dimension of their \Teich spaces, without
proofs.
 The \Teich spaces of Seifert bundles modeled on maximal geometries
are
discussed by Ohshika \cite{Oh} and Kulkarni {\it et al.}\/
\cite{KLR}.

\medskip
 A quotient $X/\Gamma$ of a subgeometry $(X,G\pr)$ of
a geometry $(X,G)$ can also be viewed as that of $(X,G)$.
 Since $\Gamma$ is a subgroup of $G\pr$ and $G\pr$ is a subgroup of
$G$,
$\Gamma$ is a discrete subgroup of $G$, which implies $X/\Gamma$ is
a quotient of $(X,G)$.

\medskip
 Thus, all compact locally homogeneous Riemannian manifold
and their \Teich spaces are obtained by the following procedure.

 1)Pick up a Thurston geometry $X$ (say, $\E3$) and
list up all compact locally homogeneous manifolds $M\cong X/\Gamma$
modeled on $(X,\Isom^+X)$.

 2)List up all types (say, Type a1 and Type a2) of universal covers
$\ti M$
which is subgeometries of $(X,\Isom^+X)$.

 3)For each universal cover type, check whether $X/\Gamma$ can be
modeled on
$(\ti M,\Isom\ti M)$ or not by checking whether $\Gamma$ is a
subgroup of
$\Isom\ti M$.

 4)Find their \Teich spaces.

 Note that a representation $\rho:\pi_1(M)\to\Isom^+\ti M$ is
determined
by the images of the fixed generators of $\pi_1(M)$.

 For the universal covers which have L and R types, we show the
results of L type; the representations in R type can be obtained in
the
same way.

\subsect{Type a}
\label{type a}
 There are following six compact orientable quotients modeled on
$\E3$,
i.e. Type a2.
(Theorem 3.5.5 of \cite{Wo}).

 1) The fundamental group is generated by $a,b$, and $c$ and the
relations are
\beq
\label{eq:a/1}
[a,b]\equiv aba^{-1}b^{-1}=1,\ [a,c]=1,\ [b,c]=1,
\eeq
 which will be denoted as
\beq
\pi_1(M)=\langle a,b,c ;[a,b],[a,c],[b,c]\rangle.
\eeq
 Here the convention is adopted that a product $ab$ denotes
a turn of curve $b$ followed by a turn of curve $a$, so that
the same expression can be used for both an element of the
fundamental group and an element of the discrete group in the
isometry group.
 When $\pi_1(M)$ is represented in $\Isom\E3$,
the generators $a,b$, and $c$ are
translations in different directions.
The manifold is a torus, $\T3$.

 2) The fundamental group is
\beq
\pi_1(M)=\langle a,b,c ;[a,b],cac^{-1}a,cbc^{-1}b\rangle.
\eeq
 In $\Isom\E3$, $a$ and $b$ are translations and $c$ is a
screw motion with a rotation angle $\pi$.
 So the direction of $c$ must be orthogonal to those of $a$
and $b$.
 The manifold is $\T3/\ZZ2$.

 3) The fundamental group is
\beq
\pi_1(M)=\langle a,b,c ;[a,b],cac^{-1}b^{-1},cbc^{-1}ba\rangle.
\eeq
 In $\Isom\E3$, $a$ and $b$ are translations and $c$ is a
screw motion with a rotation angle $2\pi/3$.
 The direction of $c$ must be orthogonal to those of $a$
and $b$.
 The vectors of translations $a$ and $b$
must have the same length.
 The manifold is $\T3/\ZZ3$.

 4) The fundamental group is
\beq
\pi_1(M)=\langle a,b,c ;[a,b],cac^{-1}b^{-1},cbc^{-1}a\rangle.
\eeq
 In $\Isom\E3$, $a$ and $b$ are translations and $c$ is a
screw motion with a rotation angle $\pi/2$.
 The direction of $c$ must be orthogonal to those of $a$
and $b$.
 The vectors of translations $a$ and $b$
must have the same length.
 The manifold is $\T3/\ZZ4$.

 5)  The fundamental group is
\beq
\pi_1(M)=\langle a,b,c ;[a,b],cac^{-1}b^{-1},cbc^{-1}b^{-1}a\rangle.
\eeq
 In $\Isom\E3$, $a$ and $b$ are translations and $c$ is a
screw motion with a rotation angle $\pi/3$.
 The direction of $c$ must be orthogonal to those of $a$
and $b$.
 The vectors of translations $a$ and $b$
must have the same length.
 The manifold is $\T3/\ZZ6$.

 6)  The fundamental group is
\beqa
\pi_1(M)&=& \langle a,b,c ;cab^{-1},ab^2a^{-1}b^2,ba^2b^{-1}a^2
\rangle\nono\\
        &=&  \langle a,c; ca^2c^{-1}a^2,ac^2a^{-1}c^2 \rangle.
\eeqa
 In $\Isom\E3$, all generators are screw motions with a rotation
angle $\pi/2$.

\medskip
 Let $\Gamma_i (i=1,2,\cdots,6)$ denote the images of the
representations
of the fundamental groups from 1) to 6).
 We will write the quotient of the universal cover of Type a2 by
$\Gamma_1$
as $\ti M_{\rm a2}/\Gam_1$ or simply as a2/1, and so forth.

\medskip
 All fundamental groups have representations in $\Isom^+\ti M_{a1}$.
 The first five ones do because Type a1 has translations in a plane
and
one screw motion in the orthogonal direction to the plane.
 The last one also does by virtue of the discrete elements of
$\Isom^+\ti M_{a1}$, as we will see later.

\medskip
 Let us find the representations of $\pi_1(\ti M_{\rm a1}/\Gamma_1)$

in $\Isom\ti M_{a1}$.
  It must be faithful, discrete in $\Isom^+\ti M$, and has no fixed
points.
 First we consider the case that $\pi_1(M)$ is represented
in $\Isom_0 \ti M_{a1}$.
 Let us denote an element $a$ of $\Isom_0 \ti M_{a1}$ as
\beq
a=\left(\ba{c}
 a_1\\a_2\\a_3
\ea\right)\equiv\mbox{Exp}a_2A_2\circ\mbox{Exp}a_1A_1\circ\mbox{Exp}a_
3A_3,
\eeq
where $A_i$ are vector fields in eq. (\ref{eq:alg:vii0}).
 This is a screw motion along the $z$-axis followed
by a translation in the $xy$-plane.
 The multiplication rule for Bianchi VII(0) group is then found by
explicit calculations,
\beq
\label{eq:vii0:multiplication}
ab=\left(\ba{c}a_1\\a_2\\a_3\ea\right)
\left(\ba{c}b_1\\b_2\\b_3\ea\right)\nono\\
=\left(\ba{c}
\left(\ba{c}a_1\\a_2\ea\right) +R_{a_3}
\left(\ba{c}b_1\\b_2\ea\right)\\
a_3+b_3
\ea\right).
\eeq
 The inverse of $a$ is
\beq
a^{-1}=
\left(\ba{c}-R_{-a_3} \left(\ba{c}a_1\\a_2\ea\right) \\-a_3
\ea\right).
\eeq

 The representation of the $\pi_1(\ti M_{\rm a1}/\Gam_1)$ is obtained
by
writing generators with components and demanding (\ref{eq:a/1}).
 The solution is
\beq
\label{eq:rep:a/1:1}
a=\left(\ba{c} a_1\\a_2\\2l\pi\ea\right),
b=\left(\ba{c} b_1\\b_2\\2m\pi\ea\right),
c=\left(\ba{c} c_1\\c_2\\2n\pi\ea\right)
\eeq
where $l,m$ and $n$ are integers.

 Similarly, the representations including elements of $\Isom\ti M$
not connected to the
identity are generated by $a,b$ and $c=h\circ c\pr$ where
$h:(x,y,z)\mapsto(-x,-y,z)$ and
 \beq
\label{eq:rep:a/1:2}
a=\left(\ba{c} a_1\\a_2\\2l\pi\ea\right),
b=\left(\ba{c} b_1\\b_2\\2m\pi\ea\right),
c\pr=\left(\ba{c} c_1\\c_2\\(2n+1)\pi\ea\right)
\eeq
 where we have used that $h\circ c\circ h^{-1}=h(c)=(-c_1,-c_2,c_3)$.

\medskip
 The \Teich space is obtained by taking conjugacy class of discrete
groups by
$\Isom_0\ti M=$Bianchi VII(0) group, for the universal cover has no
global conformal isometries.
 In the case that the generators of $\Gam$ is (\ref{eq:rep:a/1:1}),
the conjugate of the generators by
$s=\left(\ba{c} s_1\\s_2\\s_3\ea\right) \in\Isom_0\ti M_{a1}$
 is
\beq
sas^{-1}=
\left(\ba{c}R_{s_3} \left(\ba{c}a_1\\a_2\ea\right)\\2l\pi
\ea\right),
sbs^{-1}=
\left(\ba{c}R_{s_3} \left(\ba{c}b_1\\b_2\ea\right)\\2m\pi
\ea\right),
scs^{-1}=
\left(\ba{c}R_{s_3} \left(\ba{c}c_1\\c_2\ea\right)\\2n\pi
\ea\right).
\eeq
We see that the vectors
$\left(\ba{c}a_1\\a_2\ea\right),
\left(\ba{c}b_1\\b_2\ea\right)$ and
$\left(\ba{c}c_1\\c_2\ea\right)$
are rotated by the conjugation.
 So we can choose a representative element of the equivalence class
corresponding
to a point in the \Teich space as
\beq
\label{eq:teich:a/1}
a=\left(\ba{c}a_1\\0\\2l\pi\\\ea\right),
b=\left(\ba{c}b_1\\b_2\\2m\pi\ea\right),
c=\left(\ba{c}c_1\\c_2\\2n\pi\ea\right),
\eeq
where $a_1>0$.
 In the case of (\ref{eq:rep:a/1:2}) the representative is
\beq
a=\left(\ba{c}a_1\\0\\2l\pi\ea\right),
b=\left(\ba{c}b_1\\b_2\\2m\pi\ea\right),
c=h\circ\left(\ba{c}c_1\\c_2\\(2n+1)\pi\ea\right).
\eeq
 It follows from above that the dimension $T$ of the \Teich space is
five.

\medskip
 Along the same line of argument, we can see that
for the other fundamental groups the representative elements of
the equivalence class corresponding to points in the
\Teich space and its dimension are the following:
\begin{eqnarray}
    {\rm a1}/2:\quad a\wa\vector{a_1}{0}{0},b=\vector{b_1}{b_2}{0},
    c=\vector{0}{0}{(2n+1)\pi}; \\
    a\wa\vector{a_1}{0}{0},b=\vector{b_1}{b_2}{0},
    c=h\circ\vector{0}{0}{2n\pi}; \\
    a\wa\vector{0}{a_2}{2m\pi},b=\vector{0}{b_2}{2n\pi},
    c=hk\circ\vector{c_1}{0}{0},\, (a_2>0); \\
    a\wa\vector{a_1}{0}{2m\pi},b=\vector{b_1}{0}{2n\pi},
    c=k\circ\vector{0}{c_2}{0}. \\
    T\wa\mbox{dim(Teich}(\ti M_{\rm a1}/\Gam_1))=3. \\
    {\rm a1}/3:\quad a\wa\vector{a_1}{0}{0},
    b=\svector{R_{\pm2\pi/3}\svector{a_1}{0}}{0},
    c=\vector{0}{0}{(2n\pm{2\over3})\pi}; \\
a\wa\vector{a_1}{0}{0},b=\svector{R_{\pm2\pi/3}\svector{a_1}{0}}{0},
    c=h\circ\vector{0}{0}{(2n\mp{1\over3})\pi}. \\
    T\wa1. \\
    {\rm a1}/4:\quad a\wa\vector{a_1}{0}{0},
    b=\svector{R_{\pm\pi/2}\svector{a_1}{0}}{0},
    c=\vector{0}{0}{(2n\pm{1\over2})\pi}; \\
a\wa\vector{a_1}{0}{0},b=\svector{R_{\pm\pi/2}\svector{a_1}{0}}{0},
    c=h\circ\vector{0}{0}{(2n\mp{1\over2})\pi}; \\
    T\wa1. \\
    {\rm a1}/5:\quad a\wa\vector{a_1}{0}{0},
    b=\svector{R_{\pm\pi/3}\svector{a_1}{0}}{0},
    c=\vector{0}{0}{(2n\pm{1\over3})\pi}; \\
a\wa\vector{a_1}{0}{0},b=\svector{R_{\pm\pi/3}\svector{a_1}{0}}{0},
    c=h\circ\vector{0}{0}{(2n\mp{2\over3})\pi}; \\
    T\wa1. \\
    {\rm a1}/6:\quad a\wa hk\circ\vector{a_1}{0}{2n\pi},
    c=k\circ\vector{0}{c_2}{0};{\rm etc.} \\
    T\wa2.
\end{eqnarray}
 Here $k:(x,y,z)\mapsto(-x,y,-z),\;a_1>0$ and the double signs are
taken
in the same order.
 Most of the other parameters should not vanish for
the representations to be
faithful.
 We have used the fact that $h\circ a\circ h=h(a)=(-a_1,-a_2,a_3)$
and
$k\circ a\circ k=k(a)=(-a_1,a_2,-a_3)$ holds for any $a\in\Isom\ti
M_{a1}$,
which can be easily verified by the explicit calculation of the
action of $h,k$ and $a$ on $\ti M$.

\medskip
 The dimension of the \Teich spaces coincide with those of Type a2.
 One can verify that the \Teich parameters of Type a1 has
corresponding
ones of Type a2.

\subsect{Type b}
 There are seven types of compact orientable quotients modeled on
$\Nil$ i.e.
Type b; each type consists of infinitely many manifolds.
 Any quotient admits a Seifert bundle structure over a Euclidean
orbifold.

 1)$\S1$-bundle over $\T2$.
\beq
\pi_1(M)=\langle a,b,c ;[a,b]c^{-n},[a,c],[b,c]\rangle.
\eeq
 Generators $a$ and $b$ correspond to the nontrivial loops of the
base $\T2$
and $c$ corresponds to that of the fiber $\S1$.
 The first relation means that if one goes through the both
nontrivial
closed curves of the base one has turned the fiber $n$ times.
\beqa
& &\mbox{ 2)$\S1$-bundle over a Klein bottle.\hspace*{9.5cm}} \nono\\
& &\pi_1(M)=\langle a,b,c ;abab^{-1}c^{-n},[a,c],bcb^{-1}c\rangle.\\
& &\mbox{ 3)}
\pi_1(M)=\langle a,b,c ;[a,b]c^{-2n},cac^{-1}a,cbc^{-1}b\rangle.\\
& &\mbox{ 4)}
\pi_1(M)=\langle a,b,c
;[a,b]c^{-3n},cac^{-1}b^{-1},cbc^{-1}ba\rangle.\\
& &\mbox{ 5)}
\pi_1(M)=\langle a,b,c
;[a,b]c^{-4n},cac^{-1}b^{-1},cbc^{-1}a\rangle.\\
& &\mbox{ 6)}
\pi_1(M)=\langle a,b,c ;
[a,b]c^{-6n},cac^{-1}b^{-1},cbc^{-1}b^{-1}a\rangle.\\
& &\mbox{ 7)}
\pi_1(M)=\langle a,c ;ac^2a^{-1}c^2,ca^2c^{-1}a^2c^{2n}\rangle.
\eeqa
\noindent
 Topologies are classified by a positive integer $n$.

\medskip
 The composition of two elements $a,b$ of $\Isom_0\ti M_{\rm b}$ are
\beqa
ab&=&\left( \ba{c}a_0\\a_1\\a_2\\a_3 \ea\right)
\left( \ba{c}b_0\\b_1\\b_2\\b_3 \ea\right)\nono\\
&=&\left( \ba{c}a_0+b_0\\
\left(\ba{c}a_1\\a_2\ea\right)+R_{a_0}
\left(\ba{c}b_1\\b_2\ea\right)\\
a_3+b_3+a_1(b_1\sin a_0+b_2\cos a_0)
+{1\over2}((b_1^2-b_2^2)\cos a_0-2b_1 b_2\sin a_0)\sin a_0\ea
\right)\nono\\
& &\ \ \ \
\eeqa
where
\beqa
\left( \ba{c}a_0\\a_1\\a_2\\a_3 \ea\right)
&\equiv &\mbox{Exp}a_3A_3\circ\mbox{Exp}a_2A_2
   \circ\mbox{Exp}a_1A_1\circ\mbox{Exp}a_0A_0,\\
& &A_1\equiv \dif x+y\dif z, A_2\equiv \dif y, A_3\equiv\dif z,\nono\\
& &A_0\equiv\dif \theta=-y\dif x+x\dif y+{1\over2}(x^2-y^2)\dif z.
\eeqa
 This can be verified by writing $a\in\Isom_0\ti M_{\rm b}$ as
\bea
a=\left( \ba{c}0\\0\\0\\a_3 \ea\right)
\left( \ba{c}0\\a_1\\a_2\\0 \ea\right)
\left( \ba{c}a_0\\0\\0\\0 \ea\right)
\eea
and use the fact that $\{\mbox{Exp}tA_3\;|\;t\in\R\}$  is the center
of $\Isom_0\ti M_{\rm b}$ and that
$s_\theta\circ q\circ s_\theta^{-1}=s_\theta(q)$
where $s_\theta$ and $q$ are isometries and
$s_\theta=\mbox{Exp}\theta A_0$ and $q\in\Nil$.

\medskip
 Representative elements of equivalence classes corresponding to
points
in \mbox{Teich}$(M)$ are given by
\bea
\mbox{b/1:     }
& &a=\left(\ba{c}0\\a_1\\0\\0\ea\right),
b=\left(\ba{c}0\\b_1\\b_2\\0\ea\right),
c=\left(\ba{c}0\\0\\0\\{1\over n}a_1b_2\ea\right).\\
& &T=3.\\
\mbox{b/2:     }
& &a=\left(\ba{c}0\\a_1\\0\\0\ea\right),
b=h\circ\left(\ba{c}0\\0\\b_2\\0\ea\right),
c=\left(\ba{c}0\\0\\0\\{1\over n}a_1b_2\ea\right).\\
& &T=2.\\
\mbox{b/3:     }
& &a=\left(\ba{c}0\\a_1\\0\\0\ea\right),
b=\left(\ba{c}0\\b_1\\b_2\\0\ea\right),
c=\left(\ba{c}\pi\\-b_1\\0\\{1\over2n}a_1b_2\ea\right).\\
& &T=3.\\
\mbox{b/4:     }
& &a=\left(\ba{c}0\\a_1\\0\\0\ea\right),
b=\left(\ba{c}0\\R_{\pm2\pi/3}\left(\ba{c}a_1\\0\ea\right)\\0\ea\right
),
c=\left(\ba{c}\pm{2\over3}\pi\\{1\over8}a_1\\\pm{\sqrt{3}\over8}a_1\\
\pm{\sqrt{3}\over2}\left({1\over3n}+{1\over192}\right)a_1^2\ea\right).
\\
& &T=1.\\
\mbox{b/5:     }
& &a=\left(\ba{c}0\\a_1\\0\\0\ea\right),
b=\left(\ba{c}0\\0\\\pm a_1\\0\ea\right),
c=\left(\ba{c}\pm{\pi\over2}\\0\\0\\\pm{1\over4n}a_1^2\ea\right).\\
& &T=1.\\
\mbox{b/6:     }
& &a=\left(\ba{c}0\\a_1\\0\\0\ea\right),
b=\left(\ba{c}0\\R_{\pm\pi/3}\left(\ba{c}a_1\\0\ea\right)\\0\ea\right)
,
c=\left(\ba{c}\pm{\pi\over3}\\-{3\over8}a_1\\\mp{\sqrt{3}\over8}a_1\\
\pm{\sqrt{3}\over2}\left({1\over6n}-{3\over64}\right)a_1^2\ea\right).
\\
& &T=1.\\
\mbox{b/7:     }
& &a=h\circ\left(\ba{c}\pi\\a_1\\a_2\\0\ea\right),
c=\left(\ba{c}0\\0\\0\\-{1\over n}a_1a_2\ea\right).\\
& &T=2.\\
\eea
 Here $h:(x,y,z)\mapsto(-x,y,-z),a_1>0$, and $a_2,b_2\neq0$ in all
cases.
 We have used the fact that
\beq
h\circ \left(\ba{c}a_0\\a_1\\a_2\\a_3\ea\right)\circ h
=\left(\ba{c}-a_0\\-a_1\\a_2\\-a_3\ea\right)
\eeq
holds for any $a\in\Isom\ti M_{\rm b}$,
which can be easily verified by the explicit calculation of the
action
of $h$ and $a$ on $\ti M$.

\medskip
 It can be shown that the \Teich space is homeomorphic to that of
the base orbifold.
 So all \Teich parameters of quotients of Type b can be considered as
those of the base orbifolds.

\subsect{Type c}
 It is known that any compact quotient of $\H2\times\R$ i.e. of the
universal
cover of Type c admits a unique Seifert bundle structure with $e=0$
whose base is a hyperbolic orbifold.
 It comes from $e=0$ that the quotient is finitely covered by
an orientable trivial $\S1$-bundle over an orientable compact
hyperbolic 2-manifold.
 Since each quotient admit a unique Seifert bundle structure,
the topologies of the quotients are completely
classified by whether the base orbifold is
orientable or not, the genus $g$ of the base, the number $k$ of cone
points,
and the Seifert invariants
$(\alpha_i,\beta_i)(i=1,\cdots,k)$ consistent with $e=0$.

\medskip
 The dimension $T$ of the \Teich space is given by $8g-5+2k$
if the base orbifold
is orientable and $4g-6+2k$ if nonorientable.
 One degree of freedom corresponds to
the length of the circle relative to the base.
 The remainder correspond to the dimension of the \Teich space
of the base orbifold
and the twist of the fiber around each nontrivial closed curve of the
base orbifold.
In the former case the dimension of the \Teich space of the base is
$6g-6+2k$
where $2k$ corresponds to the positions of the cones on the
2-surface,
and the base has $2g$ independent nontrivial closed curves.
 In the latter case the dimension of the \Teich space of the base is
$3g-6+2k$.
 There are $g$ generators of the fundamental group of the base but
the twists
of the fiber around them have $g-1$ degrees of freedom because of
conjugacy.

\subsect{Type d}
 Similar to Type c, any compact quotient of $\tiSLR2$ i.e. of the
universal
cover of Type d admits a unique Seifert bundle structure
whose base is a hyperbolic orbifold but with $e\neq0$.
 Each quotient admit a unique Seifert bundle structure.
 The topologies are completely classified by whether the base
orbifold is
orientable, the genus $g$ of the base orbifold,
the number $k$ of cone points, and the Seifert invariants
$(\alpha_i,\beta_i)(i=1,\cdots,k)$ and the Euler number $e(\neq0)$.

\medskip
 The dimension $T$ of the \Teich space is given by $8g-6+2k$
if the base orbifold
is orientable and $4g-7+2k$ if it is nonorientable.
 This is the same as the case of Type c but the length of the fiber
must be fixed because the universal cover Type d is the unit tangent
bundle
of \SLR2.
 Note that this degree of freedom exists in the universal cover.
 In fact, universal covers which have the same isometry group can be
obtained
if we consider a tangent bundle of the constant length not equal to
one;
 the parameter $\lambda$ in Sec.\ref{univ:results:slr2}
is this degree of freedom.

\subsect{Type e}
 The geometry $(\H3, \Isom \H3)$ admits infinite number of
compact quotients.
 It is known that quotients of $(\H3, \Isom \H3)$ can be classified
by
the fundamental group.
 It has also been shown by Mostow \cite{Mo} that
each quotient does not admit \Teich
deformations.
 There are no degree of freedoms of the universal cover nor
the \Teich parameters.
 We do not investigate the variety of quotients of this type further
because the problem is not completely solved and we have
already known the degrees of freedom
of all quotients.
 The variety of quotients is discussed by Thurston \cite{Th}.

\subsect{Type f}
 Any compact quotient of Sol i.e. Type f2 is a $\T2$-bundle over
$\S1$.
 Conversely, any $\T2$-bundle over $\S1$  with hyperbolic glueing map
admits a Sol-structure.
 The fundamental group is
\beq
\pi_1(M)=\langle a,b,c; [a,b], cac^{-1}b^{-1}, cbc^{-1}ab^{-n}
\rangle,
\eeq
i.e. it is generated by $a,b$ and $c$ which obey
relations
\beqa
\label{eq:sol:pi1:1}& &[a,b]=1,\\
\label{eq:sol:pi1:2}& &cac^{-1}=b,\\
\label{eq:sol:pi1:3}& &cbc^{-1}=a^{-1}b^n.
\eeqa
 Here $c$ is the generator of the fundamental group of the base
$\S1$,
while $a$ and $b$ are those of the fiber $\T2$.
 The equations (\ref{eq:sol:pi1:2}) and (\ref{eq:sol:pi1:3})
indicate that when one goes around the base $\S1$ the fiber $\T2$ is
identified by a modular transformation.
 The topologies are classified by an integer $n$ satisfying $|n|>2$.
 Note that if $n=-1,0$ or 1 the fundamental group can be realized in
$\Isom\ti M_{\rm a1}$ ($\Isom\ti M_{\rm a2}$);
in fact they coincide with the fundamental groups
of quotients a1/3, a1/4 and a1/5 (a2/3, a2/4 and a2/5) in
Sec.\ref{type a},
respectively.

\medskip
 All representations of $\pi_1\ti M$ in $\Isom^+\ti M_{\rm f2}$ are
in $\Isom^+\ti M_{\rm f1}$.
 If $n$ is positive the representations are in
 $\Isom_0\ti M_{\rm f1}=\Sol$.
 Expressing generators with their components in $\Sol$ as
\beq
a=\left(\ba{c}a_1\\a_2\\a_3\ea\right),
b=\left(\ba{c}b_1\\b_2\\b_3\ea\right),
c=\left(\ba{c}c_1\\c_2\\c_3\ea\right)
\eeq
and putting them into (\ref{eq:sol:pi1:2}) and (\ref{eq:sol:pi1:3}),
we have
\beqa
& &a_3=b_3=0,\\
\label{eq:sol:eigenvalue}
& &\left(\ba{cc}a_1&a_2\\b_1&b_2\ea\right)
\left(\ba{ll}e^{-c_3}&0\\0&e^{c_3}\ea\right)
=\left(\ba{cc}0&1\\-1&n\ea\right)
\left(\ba{cc}a_1&a_2\\b_1&b_2\ea\right).
\eeqa
 Then (\ref{eq:sol:pi1:1}) is trivial.
 The equation (\ref{eq:sol:eigenvalue}) is an eigenvalue equation.
 The numbers $e^{-c_3}$ and $e^{c_3}$ are eigenvalues of
$\left(\ba{cc}0&1\\-1&n\ea\right)$,
i.e.
\beq
c_3=\mbox{ln}{{n+\sqrt{n^2-4}}\over2}
\eeq
and
\beq
\left(\ba{c}a_1\\b_1\ea\right)=\alpha\left(\ba{c}u_1\\v_1\ea\right),
\left(\ba{c}a_2\\b_2\ea\right)=\beta\left(\ba{cc}u_2\\v_2\ea\right),
\eeq
where $\left(\ba{c}u_1\\v_1\ea\right)$ and
$\left(\ba{c}u_2\\v_2\ea\right)$
are the normalized eigenvectors corresponding to the eigenvalues
$e^{-c_3}$ and $e^{c_3}$, respectively.
 The generators are
\bea
a=\left(\ba{c}\alpha u_1\\\beta u_2\\0\ea\right),
b=\left(\ba{c}\alpha v_1\\\beta v_2\\0\ea\right),
c=\left(\ba{c}c_1\\c_2\\c_3\ea\right).
\eea
where $\alpha$ and $\beta$ are positive parameters.

\medskip
 If $n$ is negative the representations are generated by
\bea
a=\left(\ba{c}\alpha u_1\\\beta u_2\\0\ea\right),
b=\left(\ba{c}\alpha v_1\\\beta v_2\\0\ea\right),
c=h\circ\left(\ba{c}c_1\\c_2\\c_3\ea\right),
\eea
where $-e^{-c_3}$ and $-e^{c_3}$ are eigenvalues of
$\left(\ba{cc}0&1\\-1&n\ea\right)$;
$\left(\ba{c}u_1\\v_1\ea\right)$ and $\left(\ba{c}u_2\\v_2\ea\right)$
are corresponding unit eigenvectors;
and $h:(x,y,z)\mapsto(-x,-y,z)$.

\medskip
 Representative elements of the equivalence classes
corresponding to points in \Teich space
are obtained by taking conjugate of representations by $\Isom_0\ti
M_{\rm f1}=\Sol$.
 One finds that
\beq
a=\left(\ba{c}\alpha u_1\\\alpha u_2\\0\ea\right),
b=\left(\ba{c}\alpha v_1\\\alpha v_2\\0\ea\right),
c=\left(\ba{c}0\\0\\c_3\ea\right),
\eeq
for $n>0$ and
\beq
a=\left(\ba{c}\alpha u_1\\\alpha u_2\\0\ea\right),
b=\left(\ba{c}\alpha v_1\\\alpha v_2\\0\ea\right),
c=h\circ\left(\ba{c}0\\0\\c_3\ea\right),
\eeq
for $n<0$.
 The \Teich spaces are 1-dimensional and is parametrized by $\alpha$.

 It can be regarded as the size of the torus relative to the length
of
the circle.

\subsect{Type g}
 There are infinite number of compact quotients of $\S3$ i.e. Type g3,
all of which have finite fundamental groups.
 They do not admit any \Teich deformations, i.e. $T=0$ for all
quotients.
 It is known that any freely acting discrete subgroup of
$\Isom\ti M_{g3}$, i.e. \SO4 in $\Isom\ti M_{g2}$, i.e.
$(\SU2\times\U1)/\ZZ2$
(Theorem 4.10 of \cite{Sc}).

 The quotients $\S3/\Gamma$
which can be modeled on $(\ti M, \Isom\ti M_{g1})$ are as below,
where $n$ is a positive integer, $(q_1,q_2)$ denotes an element of
$\SU2\times\SU2$, and $\ZZ2$ in the expression of $\Gamma$
is $\{(1,1),(-1,-1)\}$.

 1)Lens spaces $L(4n,1)$.
\beqa
\Gamma &=& \{(e^{2\pi\chi_3(m/4n)},\pm
1)\;|\;m=0,\cdots,4n-1\}\;/\,\ZZ2\nono\\
       &\simeq& \ZZ{4n}.
\eeqa

 2)Lens spaces $L(2(2n-1),1)$.
\beqa
\Gamma&=&\{(e^{2\pi\chi_3(m/2(2n-1))},\pm
1)\;|\;m=0,\cdots,2(2n-1)-1\}
\;/\,\ZZ2 \nono\\
      &\simeq&\ZZ{2(2n-1)}.
\eeqa
 The manifold $\P3$ corresponds to the case of $n=1$.

 3)Lens spaces $L(2n-1,1)$.
\beqa
\Gamma&=&\{(e^{2\pi\chi_3(1/2(2n-1))},-1)^m\;|\;m=0,\cdots,2(2n-1)-1\}
\;
/\,\ZZ2 \nono\\
      &\simeq&\ZZ{2n-1}.
\eeqa
 The manifold $\S3$ corresponds to the case of $n=1$.

 4)Lens spaces $L(4(2n-1),2(2n-1)+1)$.
\beqa
\Gamma&=&\{(e^{2\pi\chi_3(m/2(2n-1))},q_2)\;|\;m=0,\cdots,2(2n-1)-1;\;
            q_2=\pm1,\pm\chi_3\}\;/\,\ZZ2 \nono\\
      &=&\{(e^{2\pi\chi_3(1/(2n-1))},\chi_3)^m,
            (-e^{2\pi\chi_3(1/(2n-1))},-\chi_3)^m\;|
m=0,\cdots,4(2n-1)-1;\}\;/\,\ZZ2 \nono\\
      &\simeq&\ZZ{4(2n-1)}.
\eeqa

 5)Prism manifolds $L(4(2n-1),2(2n-1)+1)/\ZZ2$.
\beqa
\Gamma&=&\{(e^{2\pi\chi_3(m/2(2n-1))},q_2)\;|
m=0,\cdots,2(2n-1)-1;\;
           q_2=\pm1,\pm\chi_i\;\;(i=1,2,3)\}\;/\,\ZZ2 \nono\\
      &\simeq&\ZZ{2n-1}\times D_2^*.
\eeqa

 6)$\S3/\Gamma$ with
\beq
\Gamma=(T^*\times\{\pm1\})/\ZZ2\simeq T^*.
\eeq

 7)$\S3/\Gamma$ with
\beq
\Gamma=(O^*\times\{\pm1\})/\ZZ2\simeq O^*.
\eeq

 8)$\S3/\Gamma$ with
\beq
\Gamma=(I^*\times\{\pm1\})/\ZZ2\simeq I^*.
\eeq

Here $D_n^*,T^*,O^*$ and $I^*$ denote the preimages
of $D_n,T,O$ and $I$ by the homomorphism $\SU2\to\SO3$,
where $D_n,T,O$ and $I$ denote the dihedral, tetrahedral, octahedral,
icosahedral groups, respectively \cite{Wo}.

 The lens spaces and the prism manifolds of the above types
have two degrees of freedom of the universal cover
and zero \Teich parameter, so that the total degrees of freedom is
two.

\medskip
 The other quotients are modeled on $(\ti M, \Isom\ti M_{g2})$
but not on $(\ti M, \Isom\ti M_{g1})$.
 These have one degree of universal cover
and zero \Teich parameter,
so that the total degrees of freedom is one.

\medskip
 We do not investigate the variety of the finite subgroups of
$\Isom\ti M_{g2}$ because we have already known the degrees of
freedom.
 The variety of quotients is discussed in \cite{Wo}.

\subsect{Type h}
 It is known that there exist only two compact, orientable manifold
modeled on $\S2 \times \R$ i.e. Type h.

 1) The fundamental group is isomorphic to $\Z$ i.e.
\beq
\pi_1 (M)=\langle a\rangle.
\eeq
 The manifold is $\S2 \times\S1$.

 2) The fundamental group is
\beq
\pi_1 (M)=\langle a,b;a^2,b^2\rangle.
\eeq
 The manifold is $\P3\#\P3$, which can be regarded as an
$\S1$-bundle over $\P2$.

\medskip
 The representation of $\pi_1 (M_{\rm h/1})$ in $\Isom^+\ti M_{\rm
h}$
is generated by
\beq
\label{eq:rep:h/1}
a=(R,t)
\eeq
where $R\in\SO3$ and $t$ is a translation on \R.
 So Rep$(M)$ is 4-dimensional.

\medskip
 Let $R(\hat{\bf n},\theta)$ denote a rotation around axis
$\hat{\bf n}$ and with rotation angle $\theta$,
and $t_a$ denote a translation $z\mapsto z+a$.
 The conjugate of (\ref{eq:rep:h/1}) by an element of
$\Isom_0\ti M_{\rm h}$ is
\beq
(R\pr,t_{a\pr})(R(\hat{\bf n}\pr;\theta),t_a)(R\pr{}^{-1},t_{-a\pr})
=(R(R\pr\hat{\bf n}\pr;\theta),t_a),
\eeq
 which states that the rotation axis is changed by conjugation
but the rotation angle and the length of translation are not.
 Thus the representative element of the conjugacy class
corresponding to a point in Teich$(M)$ is generated by
\beq
a=(R(\hat{\bf n}_0 ,\theta),t_a)
\eeq
where $\hat{\bf n}_0$ is a fixed axis.
 The \Teich parameters are $a\in\R-\{0\}$ and $0\le\theta\le\pi$
with $\theta=0$ and $\theta=\pi$ reflector points,
which are the length of $\S1$ relative to the size of $\S2$
and the twist angle  of $\S2$ when one goes around $\S1$.
 The \Teich space is 2-dimensional.

 Let us find representations of $\pi_1 M_{\rm h/2}$ in $\Isom^+\ti
M_{\rm h}$.
 An element $a$ of $\Isom^+\ti M_{\rm h}$ is written as
\beq
a=h^p\circ(R,t_a)
\eeq
where $h$ is a combination of the antipodal map of $\S2$
and a reflection of $\R$ at 0,
and $p$ is 0 or 1.
 The number $p$ must be 1 to satisfy the relation $a^2=1$.
 Otherwise the second entry of $a^2$ would be a translation by $2a$.
 If $p=1$ then $a$ is written as
\beq
a=(-R,r_{-a/2})
\eeq
where $r_a$ denotes a reflection at a point $a\in\R$.
 Since $a^2=(R^2,1)$, the rotation angle must be 0 or $\pi$ to
satisfy the relation $a^2=1$.
 The latter case is excluded by the requirement that $a$ should
have no fixed points.
 So the representation is generated by
\beq
a=(-1,r_{-a/2}),b=(-1,r_{-b/2})
\eeq
where $a,b\in\R-\{0\}$ and $a\neq b$.

\medskip
 The representative element of the conjugacy class has generators
\beq
a=(-1,r_{-a/2}),b=(-1,r_0).
\eeq
 The \Teich parameter is $a$.
 It is the length of a nontrivial closed curve corresponding to $ab$.
 The dimension of the \Teich space is one.

\subsect{Summary of the section}
 The compact quotients and their dynamical degrees of freedom are in
Table 3. The total degrees of freedom $F$ is the sum of the degrees
of freedom $U$ of the universal cover and the dimension $T$ of the
\Teich space.

\medskip
 In the cases that the universal covers are Type a1 and Type a2,
we only show the cases of Type a1.
 This is because a quotient of Type a2 is obtained as the special
case
of a quotient of Type a1;
the universal cover
parameters
are fixed as $\lambda_+=\lambda_-=0$.

\medskip
 Similarly, in the cases of Types f and g, we choose the universal
cover
which have the largest $U$.

\sect{Conclusion and discussions}
We have worked out the classification of topologies of compact locally
homogeneous universes except for hyperbolic 3-manifolds
by studying possible
fundamental groups for each Thurston geometry.

We have shown that the total global degrees of freedom of compact
locally
homogeneous spaces consist of two parts.
One is the universal cover parameters
while the other is the Teichm\"uller parameters which parametrize
the discrete subgroup $\Gamma$ of the isometry group.
The Teichm\"uller parameters are obtained by constructing
the representation of the fundamental group in the isometry group.

Intuitively the difference of the two degrees of freedom is the
following.
Assuming the homogeneity of the Universe one can \it locally \rm
determine
the universal cover degrees of freedom by measuring all ratios of
the sectional curvatures. In a sense these are interpreted as
local anisotropies.
In order to know the Teichm\"uller parameters, on the other hand,
we probably have to send many space explorers, who are supposed to
measure
the times taken by all possible round trips around
non-trivial loops, etc.. In short, the Teichm\"ulller parameters
are global quantities.

We have not discussed the dynamics of our compact universes in the
present
work.
In a separate paper we will study the Einstein gravity for a compact
locally homogeneous universes.

Thurston conjectured that all possible topologies
of compact 3-manifolds can be classified to be connected sums
and torus sums of the prime manifolds which are given by the compact
quotients of the eight geometries \cite{Th2}.
This may lead us to a speculation that at least some aspects of
general inhomogeneous universes are understood by studying
combinations of the compact locally homogeneous spaces.

It will be also interesting to investigate quantum field theory
in the compact locally homogeneous spaces which have non-trivial
topologies. We naturally expect the Cashmir effects, etc..

We hope our present work furnishes a mathematical base on which a new
field of cosmology develops.

\section*{Acknowledgments}
\compulsoryindent
We are greatly indebted to Professors K. Ohshika and S. Kojima
for explaining us about the Teichm\"uller parameters of the Seifert
manifolds.
We also thank to H. Kodama for a nice lecture and Y. Fujiwara for
correspondence.
T. K. expresses his sincere thanks to S. Higuchi for useful
discussions.
This work is partially supported by the Grant-in-Aid for Scientific
Research of Ministry of Education, Science and Culture of Japan
(No.02640232)(A.H.).

\section*{Appendix}
\label{classb}
\compulsoryindent
 It is proved that the Bianchi Class B
geometries \cite{EM} $(\R^3,G)$ do not admit compact quotients
\cite{AS,FK}.
 It can be shown by the same argument that a geometry $(G,G)$ where
$G$ is a Lie group of any dimension
admits compact quotients
only if the trace of the structure constants vanishes.
 Let G a group manifold which admits a compact quotient $G/\Gamma$.
 The invariant 1-forms $\{\sigma^i\}$ define the volume element
\beq
\epsilon=\sigma^1\wedge\cdots\wedge\sigma^N
\eeq
on $G$, where $N=\mbox{dim}G$.
 Let $\{X_i\}$ the dual basis of $\{\sigma^i\}$.
 It is easily verified that
\beq
{\cal L}_X\epsilon=g^{ij}C^k{}_{ik}C^l{}_{jl}\epsilon
\eeq
 and
\beq
\mbox{Exp}tX\in\mbox{Aut}(G),
\eeq
 where
\beq
X=g^{ij}C^k{}_{ik}X_j
\eeq
 If $C^k{}_{ik}\neq 0$, this means that there is a diffeomorphism
from $G/\Gamma$ onto $G/\Gamma$ which changes its volume.
 This contradicts the compactness of $G/\Gamma$.

\medskip
 The case of $N=3$ states that the Bianchi Class B geometries do not
admit compact quotients.
 It should be noted that an invariant metric under
a Bianchi Class B group may have
compact quotients when the isometry group may be larger than
that Bianchi group.
 For example, a Bianchi V-invariant metric is invariant under \PSLC2
and admits compact quotients.

\medskip
 It can be shown by using the case of $N=2$ of the above theorem that
the geometry $(\R^3,\tiSLR2)$, which belongs to the Bianchi Class A,
does not admit compact quotients.
 The case of $N=2$
of the above theorem
states that the group $\H2$ does not admit
compact quotients,
 where $\H2$ is a simply connected 2-dimensional Lie group
defined by the following multiplication rule:
\beq
\left(\ba{c}a\\b\ea\right)\left(\ba{c}x\\y\ea\right)
=\left(\ba{c}a+bx\\by\ea\right).
\eeq
 The group $\tiSLR2$ is a semidirect product of $\R$ and $\H2$ group.
 The manifold can be considered as a \R-bundle over $\R^2$;
the base is considered as a geometry $(\R^2,\H2)$.
 Since the action of $\tiSLR2$ on $\R^3$ is fiber preserving,
any quotient of the geometry inherits the same bundle
structure.
 This implies that any compact quotient of $(\R^3,\tiSLR2)$ has a
Seifert bundle structure over a compact orbifold modeled on
$(\R^2,\H2)$.
 Since the base orbifold must be finitely covered by a compact
surface
the necessary condition for $(\R^3,\tiSLR2)$ to admit compact
quotients is that the base $(\R^2,\H2)$ admits compact quotients.
 However, this is not satisfied as we have already seen.

\bthebib{99}
\bibitem{Weinberg} Weinberg, S.: Gravitation and cosmology.
New York: Wiley 1972.
\bibitem{LL} Landau, L. D.,Lifshitz, E. M.: Classical theory
of Fields. Reading: MIT 1971.
\bibitem{Wa} Wald, R. M.: General Relativity. Chicago:
University of Chicago Press 1984.
\bibitem{Na} Nariai, H.: On some static solutions of Einstein's
gravitational field equations in a spherically symmetric case.
Sci. Rep. Tohoku Univ., I, {\bf 34}, 160-167 (1950);
Nariai, H.: On a new cosmological solution of Einstein's field
equations of gravitation. {\em ibid.},\/ {\bf 35}, 62-67 (1951)
\bibitem{KS} Kantowski, R., Sachs, R. K.:
Some spatially homogeneous anisotropic relativistic cosmological
models.
J. Math. Phys. {\bf 7}, 443-446 (1967)
\bibitem{BKL} Belinski, V. A., Khalatnikov I. M., Lifshitz, E. M.:
Oscillatory approach to a singular point in the relativistic
cosmology.
Adv. in Phys. {\bf 19}, 525-573 (1970)
\bibitem{Ryan} Ryan, M. P., Shepley, L. C.: Homogeneous Relativistic
Cosmologies. Princeton Series in Physics.
Princeton: Princeton University Press 1975
\bibitem{FS} Fang, L. Z., Sato H.:
Is the periodicity in the distribution of quasar red shifts
an evidence of multiply connected universe?
Gen. Rel. Grav. {\bf 17}, 1117-1120 (1985)
\bibitem{El} Ellis, G. F. R.: Topology and cosmology.
Gen. Rel. Grav. {\bf 2}, 7-21 (1971)
\bibitem{Geroch} Geroch, R. P.: Topology in general relativity.
J. Math. Phys. {\bf 8}, 782-786 (1967)
\bibitem{Sc} Scott, P.: The geometries of 3-manifolds.
Bull. London Math. Soc. {\bf 15}, 401-487 (1983)
\bibitem{AS} Ashtekar, A., Samuel, J.:
Bianchi cosmologies: the role of spatial topology.
Class. Quantum Grav. {\bf 8}, 2191-2215 (1991)
\bibitem{Fa} Fagundes, H. V.: Closed spaces in cosmology.
Gen. Rel. Grav. {\bf 24}, 199-217 (1992)
\bibitem{Wo} Wolf, J. A.: Spaces of constant curvature, Fifth Edition.
Wilmington: Publish or Perish 1984
\bibitem{Si} Singer, I. M.: Infinitesimally homogeneous spaces.
Comm. Pure Appl. Math.\/ {\bf 13}, 685-697 (1960)
\bibitem{Oh} Ohshika, K.:
Teichm\"uller spaces of Seifert fibered manifolds with infinite
$\pi_1$.
Topology and its Application {\bf 27}, 75-93 (1987)
\bibitem{KLR} Kulkarni, R., Lee, K. B., Raymond, F.:
Deformation spaces for Seifert manifolds.
Geometry and Topology, Lecture Notes Math. {\bf 1167}, 180-216 (1985)
\bibitem{Mo} Mostow, G. D.: Strong rigidity of locally symmetric
spaces.
Ann. of Math. Studies {\bf 78}. Princeton: Princeton University Press
1973
\bibitem{Th} Thurston, W. P.: The Geometry and Topology of
3-manifolds.
To be published by Princeton University Press.
\bibitem{Th2} Thurston, W. P.:
Three dimensional manifolds, Kleinian groups and hyperbolic geometry.
Bull. Amer. Math. Soc. {\bf 6}, 357-381 (1982)
\bibitem{EM} Ellis, G. F. R., MacCallum, M. A. H.:
A class of homogeneous cosmological models.
Commun. Math. Phys. {\bf 12},108-141 (1969)
\bibitem{FK} Fujiwara, Y., Ishihara, H., Kodama, H.: Comments on
Closed Bianchi Models. Class. Quantum Grav. {\bf 10}, 859--867 (1993)

\ethebib

\newpage
\section*{Table Captions}
\compulsoryindent
{\bf Table 1:} All equivalence classes of
\[
{\mbox{\{homogeneous metrics on a simply connected manifold which
admit compact quotients\}}
\over
\mbox{\{diffeomorphisms\}
\{global conformal transformations\}}},
\]
i.e. all universal covers of compact locally homogeneous Riemannian
manifolds.

{\bf Table 2:} All equivalence classes of
\[
{\mbox{\{homogeneous metrics on a simply connected manifold which
admit compact quotients\}}
\over
\mbox{\{orientation preserving diffeomorphisms\}
\{global conformal transformations\}}},
\]
i.e. all universal covers of orientable, compact locally
homogeneous Riemannian manifolds.

{\bf Table 3:} Degrees of freedom of compact locally homogeneous
universes.
The plus and minus signs in the Types c and d denote that
the base orbifold of the Seifert bundle is orientable and
nonorientable,
respectively.

\end{document}